\documentclass[a4paper,fleqn,usenatbib]{mnras}

\usepackage[T1]{fontenc}
\usepackage{ae,aecompl}

\usepackage{graphicx}	
\usepackage{amsmath}	
\usepackage{amssymb}	

\usepackage[utf8]{inputenc}
\usepackage{relsize}
\usepackage{wasysym}

\title[Cool White Dwarfs in the SDSS DR12]{A Study of Cool White Dwarfs in the Sloan Digital Sky Survey Data Release 12}

\author[G. Ourique et al.]{
G. Ourique,$^{1}$
A. D. Romero,$^{1}$
S. O. Kepler,$^{1}$
D. Koester,$^{2}$
L. A. Amaral$^{1}$
\\
$^{1}$Instituto de Física, Universidade Federal do Rio Grande do Sul, 91501-900 Porto-Alegre, RS, Brazil\\
$^{2}$Institut für Theoretische Physik und Astrophysik, Universität Kiel, 24098 Kiel, Germany\\
}

\date{Accepted XXX. Received YYY; in original form ZZZ}

\pubyear{2018}

\begin{document}
\label{firstpage}
\pagerange{\pageref{firstpage}--\pageref{lastpage}}
\maketitle

\begin{abstract}
	In this work we study white dwarfs where $30\,000\,\text{K} {>} \mathrm{T}_{\rm{eff}} {>} 5\,000\,\text{K}$ to compare the differences in the cooling of DAs and non-DAs and their formation channels. Our final sample is composed by nearly $13\,000$ DAs and more than $3\,000$ non-DAs that are simultaneously in the SDSS DR12 spectroscopic database and in the \textit{Gaia} survey DR2. We present the mass distribution for DAs, DBs and DCs, where it is found that the DCs are ${\sim}0.15\,\mathrm{M}_\odot$ more massive than DAs and DBs on average. Also we present the photometric effective temperature distribution for each spectral type and the distance distribution for DAs and non-DAs. In addition, we study the ratio of non-DAs to DAs as a function of effective temperature. We find that this ratio is around ${\sim}0.075$ for effective temperature above ${\sim}22\,000\,\text{K}$ and increases by a factor of five for effective temperature cooler than $15\,000\,\text{K}$. If we assume that the increase of non-DA stars between ${\sim}22\,000\,\text{K}$ to ${\sim}15\,000\,\text{K}$ is due to convective dilution, $14{\pm}3$ per cent of the DAs should turn into non-DAs to explain the observed ratio. Our determination of the mass distribution of DCs also agrees with the theory that convective dilution and mixing are more likely to occur in massive white dwarfs, which supports evolutionary models and observations suggesting that higher mass white dwarfs have thinner hydrogen layers.
\end{abstract}

\begin{keywords}
white dwarfs -- catalogues -- proper motions
\end{keywords}





\section{Introduction}\label{sec:intro}

	White dwarf stars~(WDs) are the result of the stellar evolution for stars with initial masses below $7-10.6\,\mathrm{M}_\odot$, depending on the initial metalicity~\citep{2013ApJ...765L..43I, 2015MNRAS.446.2599D, 2015ApJ...810...34W}. This correspond to $97$ per cent of the stars in the Milky Way~\citep{2008ARA&A..46..157W, 2008PASP..120.1043F, 2010A&ARv..18..471A}. WDs are abundant and long-lived objects so they convey information about all galactic populations~\citep{2001ASPC..245..328I, 2005ApJS..156...47L, 2013A&A...549A.102B}. A particular feature of WDs is that their evolution is predominantly a simple cooling, so they can be considered reliable cosmic clocks to infer the age of a wide variety of stellar populations, such as the Galactic discs and halo~\citep{1987ApJ...315L..77W, 1988A&A...193..141G, 1988Natur.333..642G, 2002MNRAS.336..971T}, and globular and open clusters~\citep{2001AJ....122.3239K, 2002ApJ...574L.155H, 2007ApJ...671..380H, 2013ApJ...763..110K, 2016MNRAS.456.3729C}.

	WDs having atmospheres dominated by hydrogen, with optical spectra dominated by strong Balmer lines, are known as DAs and correspond to ${\approx}83$ per cent of the known sample of spectroscopically confirmed WDs~\citep{2004ApJ...607..426K,2006ApJS..167...40E,2015MNRAS.446.4078K,2016MNRAS.455.3413K}. WDs with spectra not dominated by hydrogen lines are known as non-DAs and can be divided into distinct sub-classes depending on the objects effective temperature and the observed spectral lines~\citep{1983ApJ...269..253S}. Around $7$ per cent of the known spectroscopic WDs have their spectra dominated by helium lines, known as DOs if they are hot enough to show He II lines and DBs if they only show He I lines~\citep[e.g.][]{2015A&A...583A..86K}. Around $5$ per cent of known spectroscopic WDs are too cool to show absorption lines in their optical spectra. These objects are known as DCs, and usually assumed to have atmospheres dominated by helium if the effective temperature is above $5\,000\,\text{K}$, inferred from the fact that hydrogen lines should be seen at those temperatures, if hydrogen is present in the atmosphere~\citep{1997ApJS..108..339B, 2001ApJS..133..413B}. Around $1$ per cent of the known WD stars present carbon lines or bands, depending on the effective temperature. WDs with atmospheres polluted or dominated by oxygen (DS) are know, but they are not statistically significant~\citep{2010Sci...327..188G, 2016Sci...352...67K}. The class of WDs showing metal lines other than carbon or oxygen in their spectra are known as DZs. These metal polluted WDs correspond to $4$ per cent of the known WDs. The spectra of DZs usually show hydrogen or helium lines if these elements are present in the atmosphere and the effective temperature is above $5\,000\,\text{K}$ or $12\,000\,\text{K}$, respectively. The metal content in the atmosphere of DZs is understood to come from ongoing accretion, since any metal from the progenitor composition will sink towards the centre of the star due to gravitational settling on time scales shorter than the WD evolutionary time scales~\citep[e.g.][] {2011A&A...530A.114K}. In addition, the accretion of hydrogen (or water) through the accretion of planetesimals can also influence the spectral class of WDs~\citep{2017MNRAS.468..971G}.
   
	\citet{2008ApJ...672.1144T} estimated the helium to hydrogen atmosphere WD number ratio as a function of effective temperature from a model atmosphere analysis of the infrared photometric data from the 2MASS survey. Their sample comprised $340$ hydrogen-rich atmosphere WDs and $107$ helium-rich atmosphere WDs with spectroscopic effective temperatures in the range $15\, 000\,\text{K} {>}\mathrm{T}_{\rm{eff}}{>}5\,000\,\text{K}$. Their focus was to study the impact of hydrogen convective mixing. They proposed that about $15$ per cent of the DAs are transformed into non-DAs at lower temperatures due to convective mixing in thin hydrogen envelope stars, where thin envelope is defined as hydrogen layer masses below ${\approx}10^{-8} \mathrm{M}_*$, with $\mathrm{M}_*$ being the stellar mass. With a total sample of $447$ WDs in their effective temperature range, they estimated that the helium to hydrogen atmosphere WD number ratio increases from $0.25$ at $15\,000\,\text{K}$ to $0.45$ at $6\,000\,\text{K}$.

	From single stellar evolution, the hydrogen layer mass not burnt in a WD should be around ${\approx}10^{-4}\mathrm{M}_*$ for canonical stellar masses of ${\approx}0.6 \,\mathrm{M}_{\odot}$, being thinner for higher masses \citep[see][] {1971AcA....21..417P, 2010ApJ...717..183R, 2012MNRAS.420.1462R}. Values below ${\approx} 10^{-8} \mathrm{M}_*$ require an additional hydrogen-depleting mechanism acting on the external layers of the stars. Strong evidence for the existence of a hydrogen layer mass range in WDs comes from asteroseismology. For instance, \citet{2009AIPC.1170..616C}, studying a sample of 83 ZZ Ceti stars, found the surface hydrogen layer to range from $10^{-10}{<}\mathrm{M}_\mathrm{H}/\mathrm{M}_*{<}10^{-4}$, with an average value of $2.71\times 10^{-5}$. \citet{2012MNRAS.420.1462R}, using asteroseismology based on full evolutionary models, analysed a sample of 44 ZZ Ceti stars with stellar masses from  ${\sim}0.5 \,\mathrm{M}_{\odot}$ to  ${\sim}0.8 \,\mathrm{M}_{\odot}$ and also found a broad range of hydrogen-layer thickness, with a similar average for the hydrogen mass. From their sample, $20$ per cent of the stars showed a hydrogen layer thinner than predicted by single stellar evolution, and $13$ per cent showed hydrogen layers below $\mathrm{M}_\mathrm{H}/\mathrm{M}_*{=}10^{-8}$. Thus, there are several pieces of evidences for mixing and spectral evolution in WDs, imprinted in the DA and non-DA samples, that could be studied in detail if a statistically significant sample of WDs is available.
    
	In recent years, the number of spectroscopically confirmed WD stars has dramatically increased, in particular due to the Sloan Digital Sky Survey~(SDSS) project. SDSS is one of the largest photometric and spectroscopic surveys, having more than $1.0$ billion objects with observed photometry and more than $4.8$ million with spectra obtained in Data Release 14~\citep{2017arXiv170709322A}. More than $30\, 000$ spectroscopically confirmed new WD stars have been found within the SDSS until Data Release 12~\citep{2004ApJ...607..426K, 2006ApJS..167...40E, 2013ApJS..204....5K, 2015MNRAS.446.4078K, 2016MNRAS.455.3413K}.
    
	With the \textit{Gaia} Data Release 2~\citep{2018arXiv180409376L}, the parallaxes of most of the known spectroscopically identified WDs have been determined, providing a direct distance determination for all objects in our sample, and allowing a better estimate of effective temperature and gravity for cool WDs that do not show significant spectral lines.
	
In this work, we take advantage of the large sample of spectroscopically confirmed WDs with measured distances to study the spectral evolution in WDs and its possible origin. We compute the distances for an initial sample of $19\,023$ WDs cooler than $30\,000,\text{K}$ from SDSS DR12 using \textit{Gaia} parallaxes, and from their colours we estimate their photometric effective temperature. Because the synthetic colour grid used for parameter determination in this work assumes $\log g = 8.0\,\text{dex}$ for DZs and DQs~\citep{2010MmSAI..81..921K,2015A&A...583A..86K} we only estimate the stellar mass for the DAs, DBs and DCs. Furthermore, we only consider DAs, DBs and DCs with estimated mass above $0.45\,\mathrm{M}_\odot$ to avoid objects that originate from interacting binaries.
	
	Our final sample has $17\,215$ objects, with $13\,678$ DAs, $1\,528$ DBs, $1\,173$ DCs, $557$ DZs and $279$ DQs. With this sample we compute the helium to hydrogen atmosphere WD number and density ratio for WDs in the range of effective temperatures $30\, 000\,\text{K}{>} \mathrm{T}_{\rm{eff}} {>} 5\,000\,\text{K}$.
	
	This paper is organised as follows: In Section \ref{sec:datadesc} we present our data sample and classification. Section \ref{sec:pardet} is devoted to mass and photometric effective temperature determination. In Section \ref{sec:spatial} we discuss the spatial distribution and completeness while in Section \ref{sec:hehratio} we present our calculation of the helium to hydrogen atmosphere WD ratio. Discussion and conclusions are presented in Section \ref{sec:conclusion}.

\section{DATA SAMPLE}\label{sec:datadesc}

	Our analysed objects are taken from the WD catalogues based on SDSS data of \citet{2004ApJ...607..426K,2006ApJS..167...40E,2015MNRAS.446.4078K,2016MNRAS.455.3413K}. We select all objects classified as single WDs with DA, DB, DC, DZ and DQ spectral type with determined effective temperature in the range $30\, 000\,\text{K}{>} \mathrm{T}_{\rm{eff}} {>} 5\,000\,\text{K}$.

	To extend our spectroscopic sample to lower effective temperatures we include the WDs in the \citet{2014AJ....148..132M} deep proper motion catalogue. \citet{2017AJ....153...10M} classified a sample of $8\,472$ objects in the deep proper motion catalogue as possible WDs using SDSS colours.
        Note that, since the cool WDs are intrinsically faint, they must be nearby and thus should have high proper motion. Of their $8\,472$ objects, $2\,144$ have spectroscopic observations from SDSS and $1\,668$ were also found in the catalogue presented in \citet{2016MNRAS.455.3413K,2015MNRAS.446.4078K,2006ApJS..167...40E,2004ApJ...607..426K}. Using the spectra from SDSS, we determine the spectral class of the $476$ objects from \citet{2017AJ....153...10M} that were not included in the SDSS WD catalogues and found that $395$ of these objects are WDs in the effective temperature range considered in this work.
	
	As a result, our initial sample is composed of $17\,492$ DAs, $1\,766$ DBs, $1\,290$ DCs, $557$ DZs and $279$ DQs, totaling $21\,384$ WDs.

\section{PARAMETER DETERMINATION}\label{sec:pardet}

	To determine the parameters characterising each object, we consider magnitudes from filters $u$, $g$, $r$, $i$ and $z$ of the SDSS photometric catalogue~\citep{1998AJ....116.3040G,2010AJ....139.1628D} de-rredened with \citet{1998ApJ...500..525S} reddening maps and using \citet{2013MNRAS.430.2188Y} extinction coefficients in addition to \textit{Gaia} parallaxes~\citep{2018arXiv180409376L}. To take into account the asymmetric uncertainties for each parameter, the parallaxes are randomly sampled assuming that the measured values are the mean values and their uncertainties are the standard deviation of a normal distribution. The sampling is repeated $1\,000$ times for each object and then the distance is calculated from each sampled parallax. This procedure generates a Monte Carlo probability density for distances. For each star we use the median of the probability distribution as the distance and the difference of $16\%$ and $84\%$ with the median as the lower and upper uncertainty, respectively.
	
	The same Monte Carlo procedure is used to sample over the magnitudes and their respective uncertainties to compute colours, which are fitted to a synthetic colour grid from \citet{2010MmSAI..81..921K,2015A&A...583A..86K} using a least squares method. We determine the absolute magnitude of synthetic colour models using the luminosity provided by \citet{2014A&A...565A..11C} models. The least squares fitting is limited to models where the photometric distances for all filters agree with the parallax distance within one sigma. This procedure generates a Monte Carlo probability density for each parameter, i.e. effective temperature and absolute magnitudes. For DAs we used synthetic colours for hydrogen-rich, while for DBs and DCs we used helium-rich atmosphere WD models. For DQs and DZs we used synthetic colours for carbon-rich and metal-rich atmosphere WD models that assumes $\log g{=}8.0$ and have abundances within $4{<}[\text{C}/\text{He}]{<}10$ and $7{<}[\text{Z}/\text{He}]{<}10.5$, respectively. For DQs and DZs models, the SDSS $u$, $g$, $r$, $i$ and $z$ do not depend on metallicity for $\mathrm{T}_{\rm{eff}}{>} 10\,000\,\text{K}$, but there is a dependence at lower temperatures, reaching one magnitude for $u$ and $0.5$ magnitude for $z$ at $5\,000\,\text{K}$. Due to the fixed $\log g$ of our atmosphere models, we could not estimate the mass of DQs and DZs.
	
	For DAs, DBs and DCs the stellar masses are estimated from the distance and effective temperature of each object, which determines their luminosity and radii, through the mass-radius relation provided by \citet{2014A&A...565A..11C} synthetic colour model grid that assumes hydrogen layer mass $\mathrm{M}_\mathrm{H}/\mathrm{M}_*{=}-4$ and helium layer mass $\mathrm{M}_\mathrm{H}/\mathrm{M}_*{=}-2$~\citep[e.g.][]{1961ApJ...134..683H,1978A&A....64..289K,1998ApJ...494..759P, 2017MNRAS.465.2849T}.

\begin{figure}
		\includegraphics[width=\columnwidth]{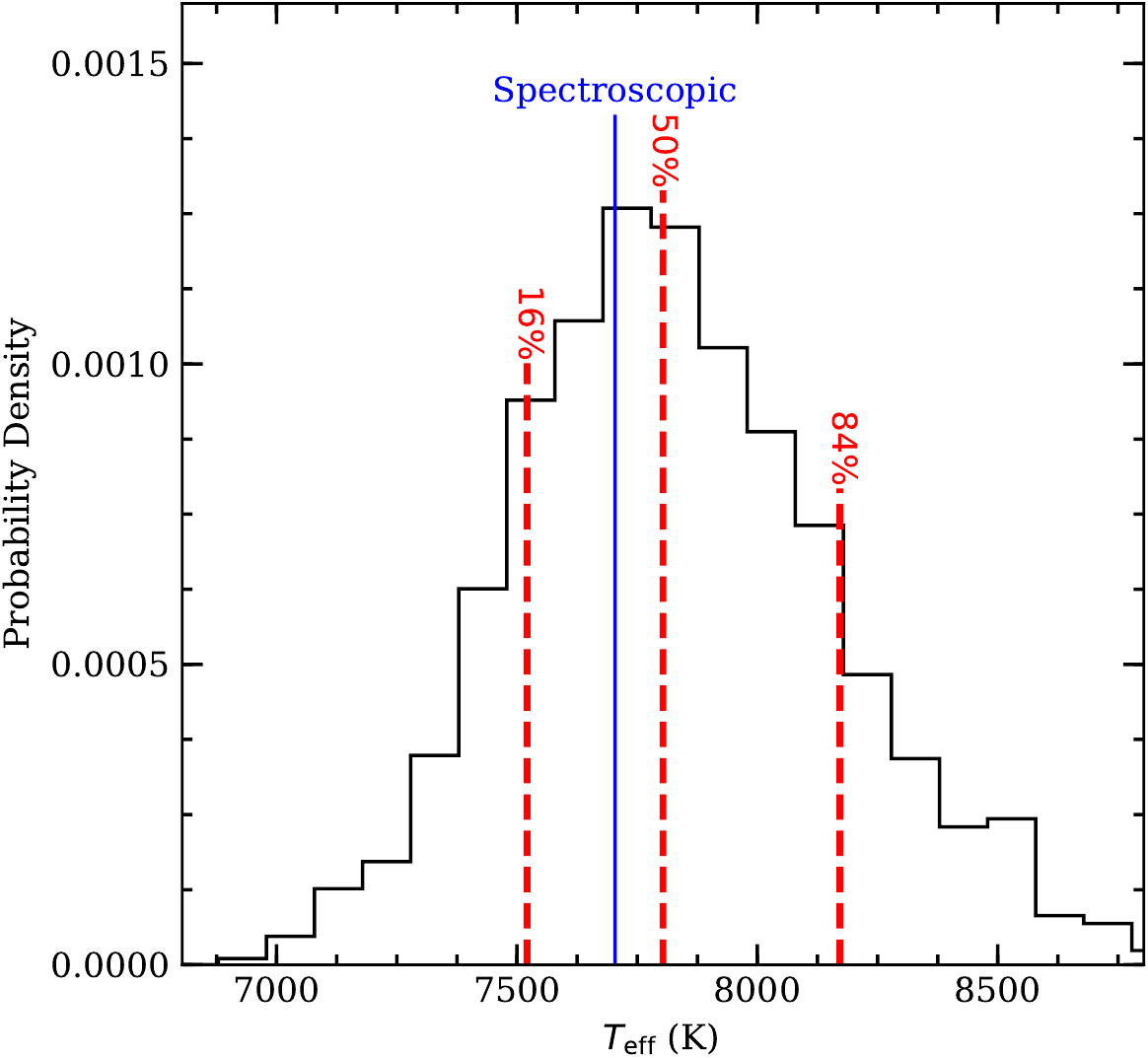}
	    	\caption{Distribution of photometric effective temperature determined with our Monte Carlo method and its uncertainty for the DAs \emph{SDSS J234212.45+044531.7}. The red dashed lines represent 16, 50 and 84 percentiles while the blue solid line is the spectroscopically determined effective temperature from \citet{2013ApJS..204....5K}. Our determination is $7804^{+368}_{-283}\,\text{K}$, agreeing within uncertainties with the spectroscopic determination, $7704\pm{56}\,\text{K}$. The SDSS magnitude in filter $g$ for this object is $19.460\pm0.019$ and the signal to noise ratio for the spectrum is $16$ in filter $g$.}
		\label{fig:Resample}
	\end{figure}

	An example of the Monte Carlo probability density for the determination of the photometric effective temperature can be seen in Figure \ref{fig:Resample} for \textit{SDSS J234212.45+044531.7}. We include in the figure, with a vertical blue line, the spectroscopic determination. The determined uncertainties are internal to our method and the photometric data, not including the uncertainties related to the synthetic colour models. 
  
	\subsection{Masses}\label{subsec:mass} 
	
\begin{figure}
		\includegraphics[width=\columnwidth]{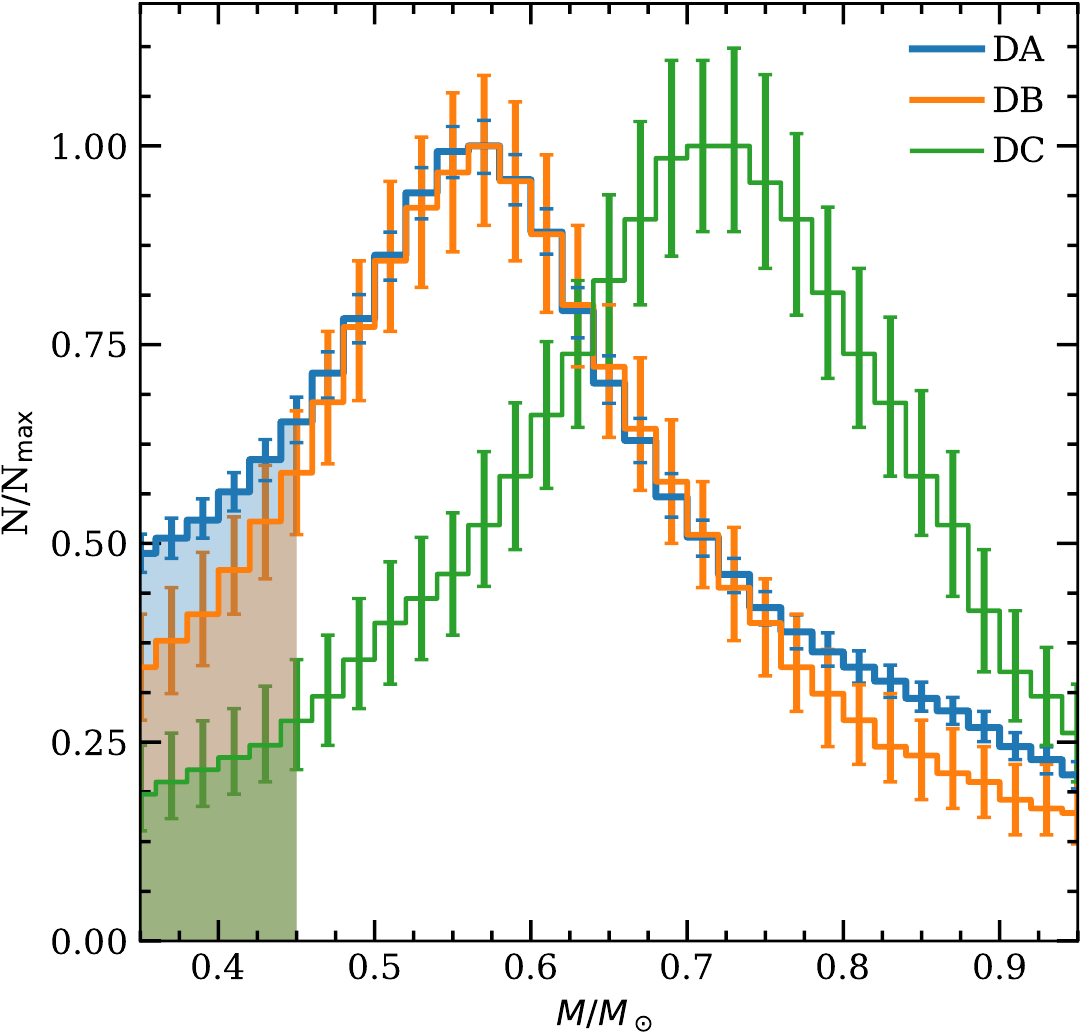}
	    	\caption{Mass distribution for DAs (blue), DBs (orange) and DCs (green). The DA and DB distributions have a peak around${\sim}0.55\,\mathrm{M}_\odot$, but the DA distribution shows higher counts in the wings of the distribution. The DCs are essentially more massive than DAs and DBs, having a peak around ${\sim}0.7\,\mathrm{M}_\odot$. The shaded regions indicate objects that originated from interacting binaries, since any object with mass below $0.45\,\mathrm{M}_\odot$ would not have time to turn into a WD by single star evolution given the age of the Universe.}
		\label{fig:distMass}
	\end{figure}

	We computed the stellar mass for all objects in our sample classified as DAs, DBs or DCs. Then we calculated the mass distribution for each spectroscopic class, shown in Figure \ref{fig:distMass}. Since we have asymmetric uncertainties that reach up to $25$ per cent of the estimated value for several objects, we used Monte Carlo methods to include this information in our distributions. Due to the large uncertainties, the study of individual objects is not reliable in this work. However, our sample is large enough for a statistically significant study. The width of our distributions come mostly from the high uncertainties on mass determinations.

	From Figure \ref{fig:distMass} we can see that the distribution in mass for DAs and DBs are very similar. The main difference between the DA and DB distributions is the higher number of counts for lower and higher masses present in the DAs mass distribution. On the other hand, the mass distribution for DCs is more concentrated towards higher masses, having a peak around ${\sim}0.7\,\mathrm{M}_\odot$, instead of ${\sim}0.55\,\mathrm{M}_\odot$ like DAs and DBs.

	Our DA sample is much larger than our DB sample and the uncertainty in mass for each object is nearly the same for both samples. We conclude that the higher counts in the wings in the DAs mass distribution are not due to uncertainties, but come from a real excess in the number of DAs at higher and lower stellar masses. Our most acceptable explanation for these objects is that they come from interacting binaries where the increase or decrease of the total mass depends on its evolution~\citep[e.g.][]{2007MNRAS.375.1315K}. This binary population is not expected in the DBs sample because binary evolution of a system containing a DB WD usually turn the helium--atmosphere component into a DA WD due to hydrogen accretion. Also a pure helium companion, or DB-DB system, is not  common enough to produce a significant number of objects~\citep[see][]{2008AstL...34..620Y}.

	Since any WD that has a stellar mass below ${\sim}0.45\,\mathrm{M}_\odot$ would take more than the age of the Universe to form through normal single star evolution~\citep{2005ApJS..156...47L,2011MNRAS.413.1121R}, we do not included objects of our sample with determined mass below $0.45\,\mathrm{M}_\odot$ to avoid binary contaminants, reducing our final sample of DAs, DBs and DCs to, $13\,678$, $1\,428$ and $1\,173$ objects, respectively.

	\subsection{Photometric Effective Temperature}
    
	In this work we determine the photometric effective temperature for our entire sample. In this way we eliminate possible effects produced by the different effective temperature determination methods or model grids used in distinct catalogues.
	
\begin{figure}
		\includegraphics[width=\columnwidth]{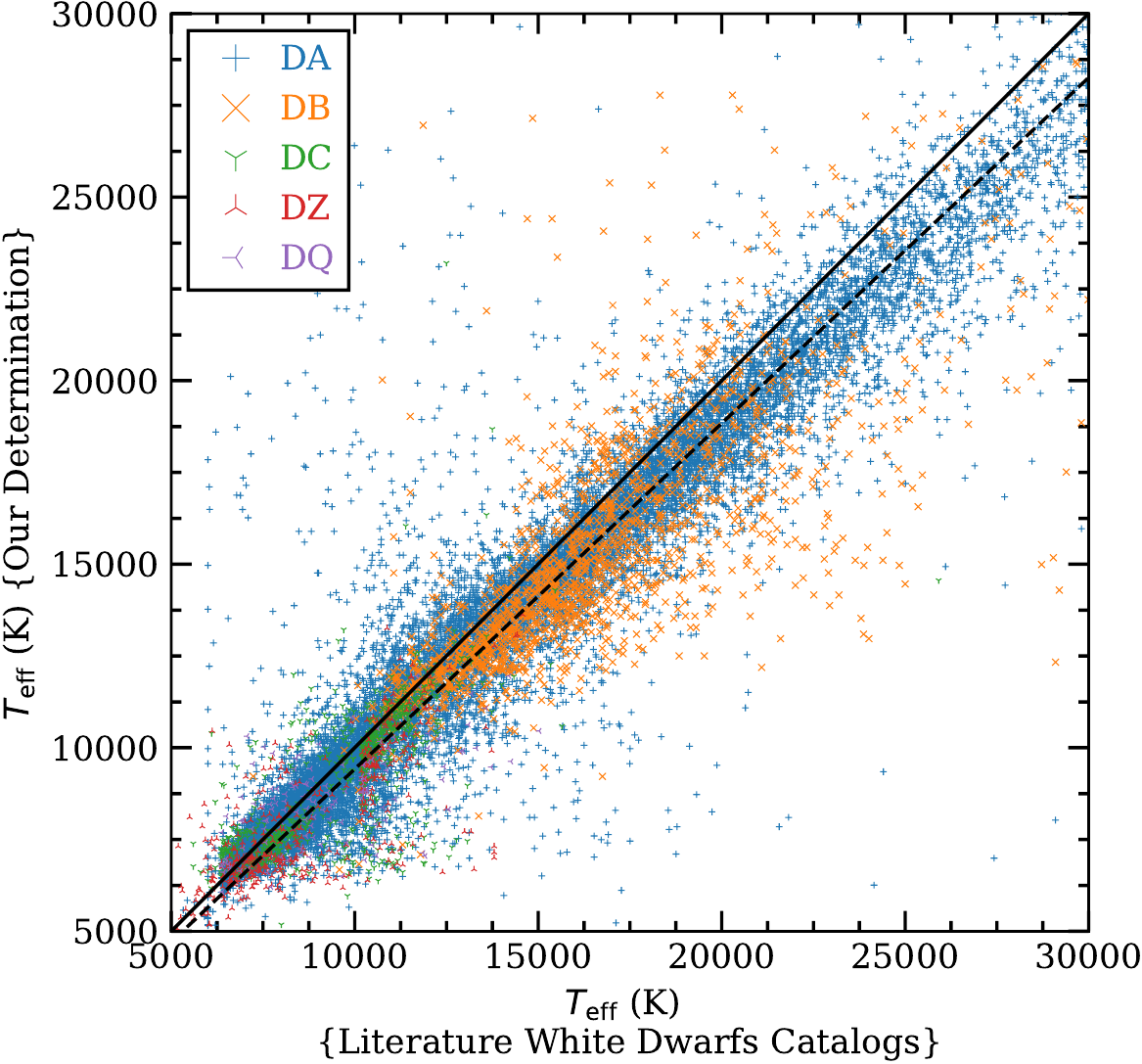}
	    	\caption{Comparison of the effective temperature determined in this work with the spectroscopic determination from \citet{2004ApJ...607..426K, 2006ApJS..167...40E, 2013ApJS..204....5K, 2015MNRAS.446.4078K, 2016MNRAS.455.3413K}. The solid black line is the identity line, while the dashed black line is the linear fit result. In colours blue, orange, green, red and purple we have the comparison for respective spectral classes: DA, DB, DC, DZ and DQ. The linear fit result indicates that our determinations are on average ${\sim}6$, ${\sim}8$, ${\sim}6$, ${\sim}10$ and ${\sim}10$ per cent cooler than the spectroscopic ones for DAs, DBs, DCs, DZs and DQs, respectively. Objects with spectroscopic determinations near to their spectroscopic model grid limit are not included.}
		\label{fig:CompTempSDSS}
	\end{figure}

	We compare our determinations of the effective temperature with the spectroscopic values from \citet{2004ApJ...607..426K, 2006ApJS..167...40E, 2013ApJS..204....5K, 2015MNRAS.446.4078K, 2016MNRAS.455.3413K} and with the photometric determination of \citet{2017AJ....153...10M}. The results are shown in Figure \ref{fig:CompTempSDSS}. Objects with spectroscopic effective temperature in the border of their spectroscopic models grid are not included in this figure. In the Figure \ref{fig:CompTempSDSS} is also plotted the identity (solid black line) and the linear fit result (dashed black line), indicating that our determinations are on average ${\approx}6$ per cent cooler than the determinations from spectroscopy. The large difference in the determination of effective temperature for bright WDs is expected due the lack of spectral energy distribution information for shorter wavelengths when we use only SDSS photometric filters. These differences are not relevant for our DA to non-DA ratio determination, since they effect both samples similarly.

\begin{figure}
		\includegraphics[width=\columnwidth]{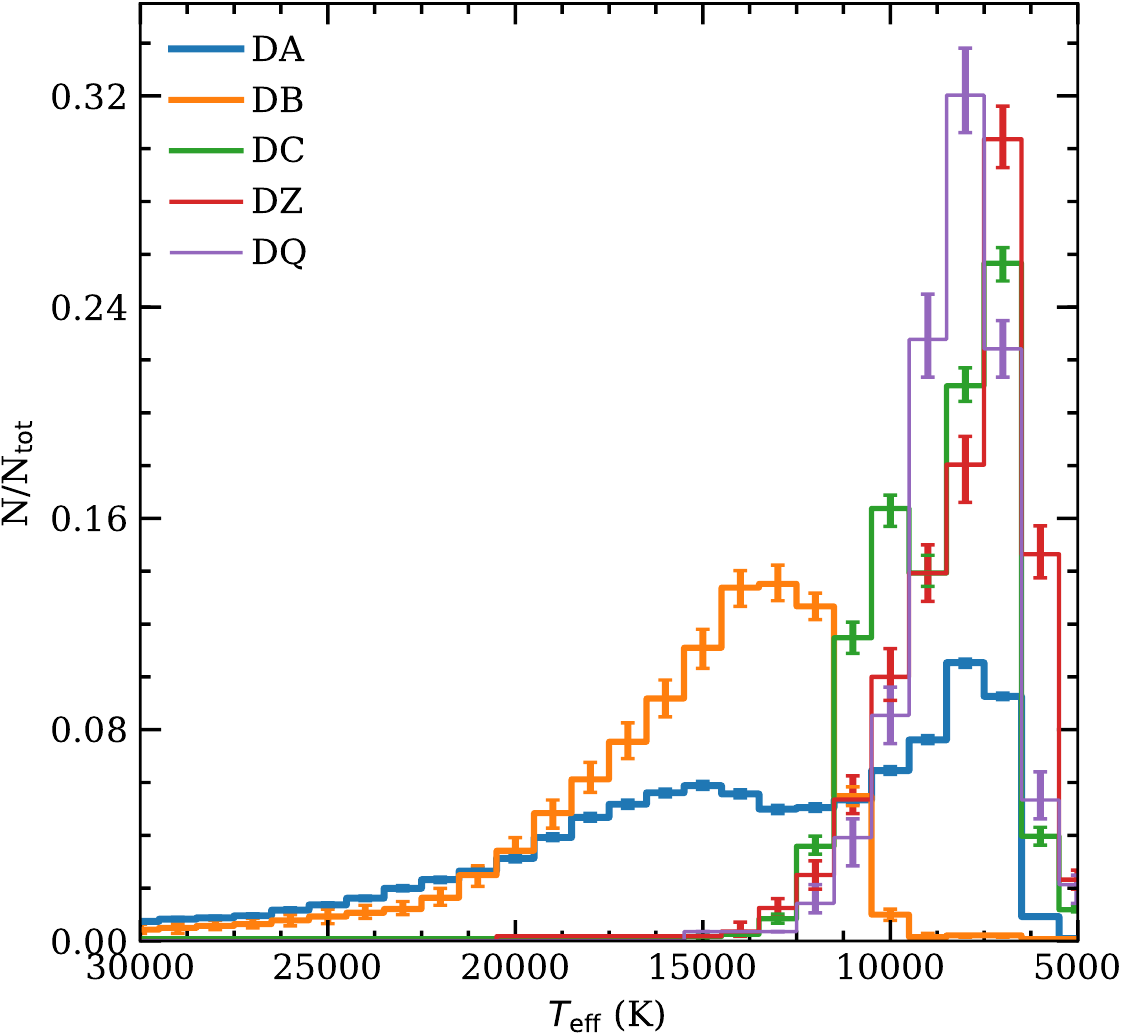}
	    	\caption{Our distribution of photometric effective temperature for DAs (blue), DBs (orange), DCs (green), DZs (red) and DQs (purple) in our sample. The bin width of all our distributions is $1\,000\,\text{K}$ and the effective temperature on the figure increase from right to left.}
		\label{fig:distTeff}
	\end{figure}

	The distribution in photometric effective temperature for DAs, DBs, DCs, DZs and DQs normalised by the number of objects in each class can be seen in Figure \ref{fig:distTeff}. Note that for DCs, DZs and DQs most objects are found at effective temperatures below $11\,000\, \text{K}$, while DBs are found above this effective temperature. On the other hand, the DAs are found in the entire effective temperature range.

	The increase in the number of WDs at lower temperatures is expected because the cooling rate decreases as they cool. The times spent at low temperatures is therefore comparatively long.
	Due to their lower luminosities, a lower completeness in the number of WDs is expected at lower effective temperatures. For DBs, the peak occurs at higher effective temperatures because DBs turn into DCs at effective temperatures lower than $12\,000\,\text{K}$ as the He lines become undetectable at signal to noise ratio ${\sim}10$.
    
	In this work, we consider as helium-rich atmosphere WDs all stars classified as DB, DC and DZ with effective temperatures in the range $30\, 000\,\text{K}{>}\mathrm{T}_{\rm{eff}}{>}5\,000\,\text{K}$, since objects with hydrogen in their atmospheres should present hydrogen lines within this temperature range~\citep[e.g.][]{2011A&A...530A.114K}. DQs are also considered as helium-rich atmosphere WDs because they do not present hydrogen lines and the carbon content is probably due to dredge-up processes~\citep{1982A&A...116..147K, 1986ApJ...307..242P, 1998MNRAS.296..523M, 2007ASPC..372...19B, 2005ASPC..334..197D}. Therefore we define as helium-rich atmosphere WDs all non-DAs from our sample. With these considerations, we study a final sample with $12\,811$ DAs and $3\,374$ non-DAs.

\begin{figure}
		\includegraphics[width=\columnwidth]{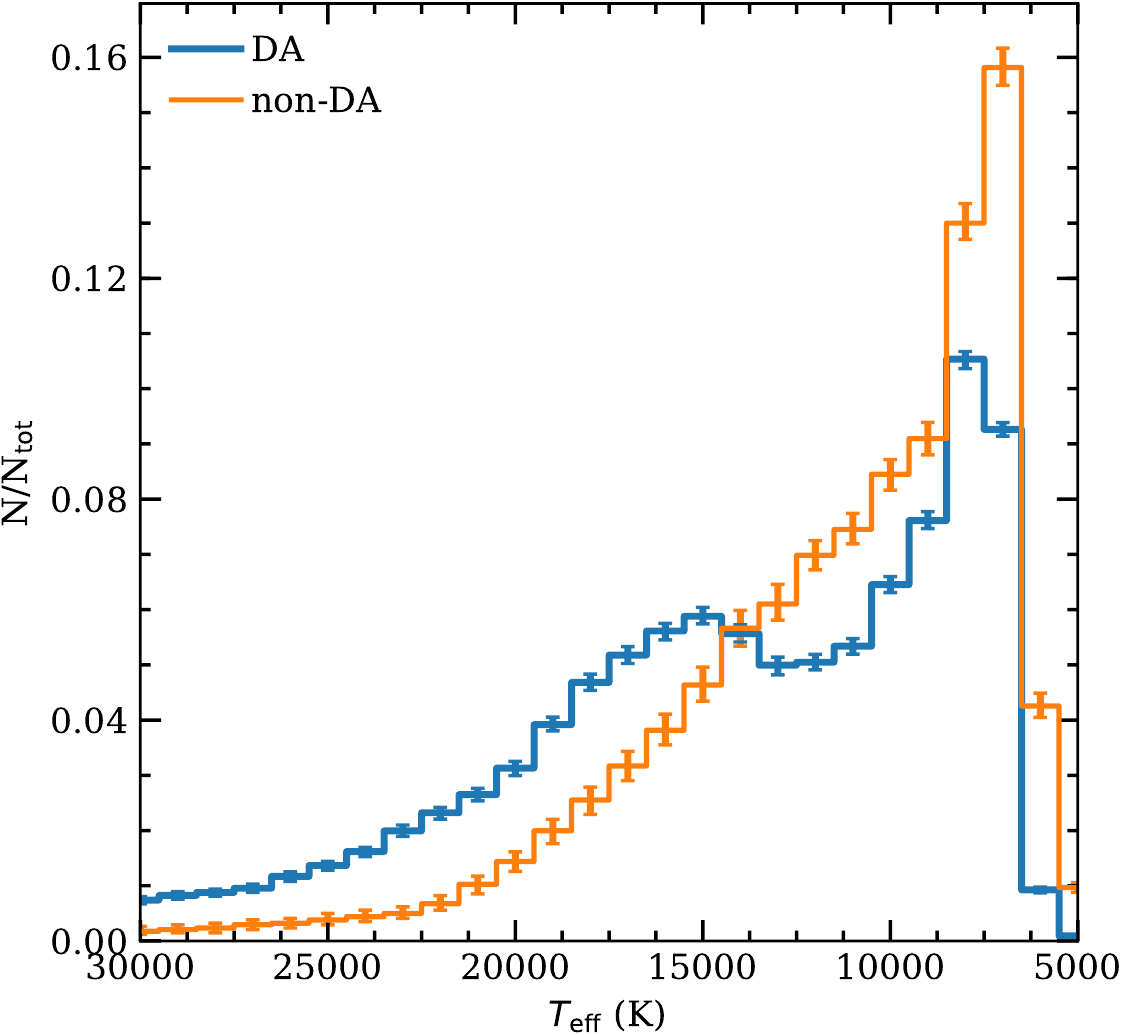}
	    	\caption{Photometric effective temperature distribution for DAs (blue) and non-DAs (orange). A change in the slope of the non-DA distribution can be seen around $21\,000\,\text{K}$, with the number of stars increasing faster for lower temperatures. The DA distribution presents a lack of stars between $14\,000\,\text{K}$ and $12\,000\,\text{K}$. The decrease in the number of counts for both distributions near to $5\,000\,\text{K}$ are due the incompleteness of the sample and the age of the disk. Both distributions are normalised by the total number of counts.}
		\label{fig:NDANDB}
	\end{figure}

	The photometric effective temperature distributions for DAs and non-DAs are depicted in Figure \ref{fig:NDANDB}, normalised by the highest number of counts. From this figure we note that the non-DA distribution shows a change in the slope around $21\, 000\,\text{K}$, making the number of counts increase faster until ${\sim}7\,000\,\text{K}$, where there is a peak in the distribution. On the other hand, the DA distribution show a decrease in the slope around $15\,000\,\text{K}$ followed by a sudden decrease between $14\,000\,\text{K}$ and $12\,000\,\text{K}$, and then increasing until ${\sim}7\,000\,\text{K}$. Both distributions show a decrease around $5\,000\,\text{K}$ due to the incompleteness of the sample and the age of the disk. Our uncertainties on the distributions are calculated using Monte Carlo methods.

\section{DISTANCE DISTRIBUTION AND COMPLETENESS}\label{sec:spatial}

\begin{figure}
		\includegraphics[width=\columnwidth]{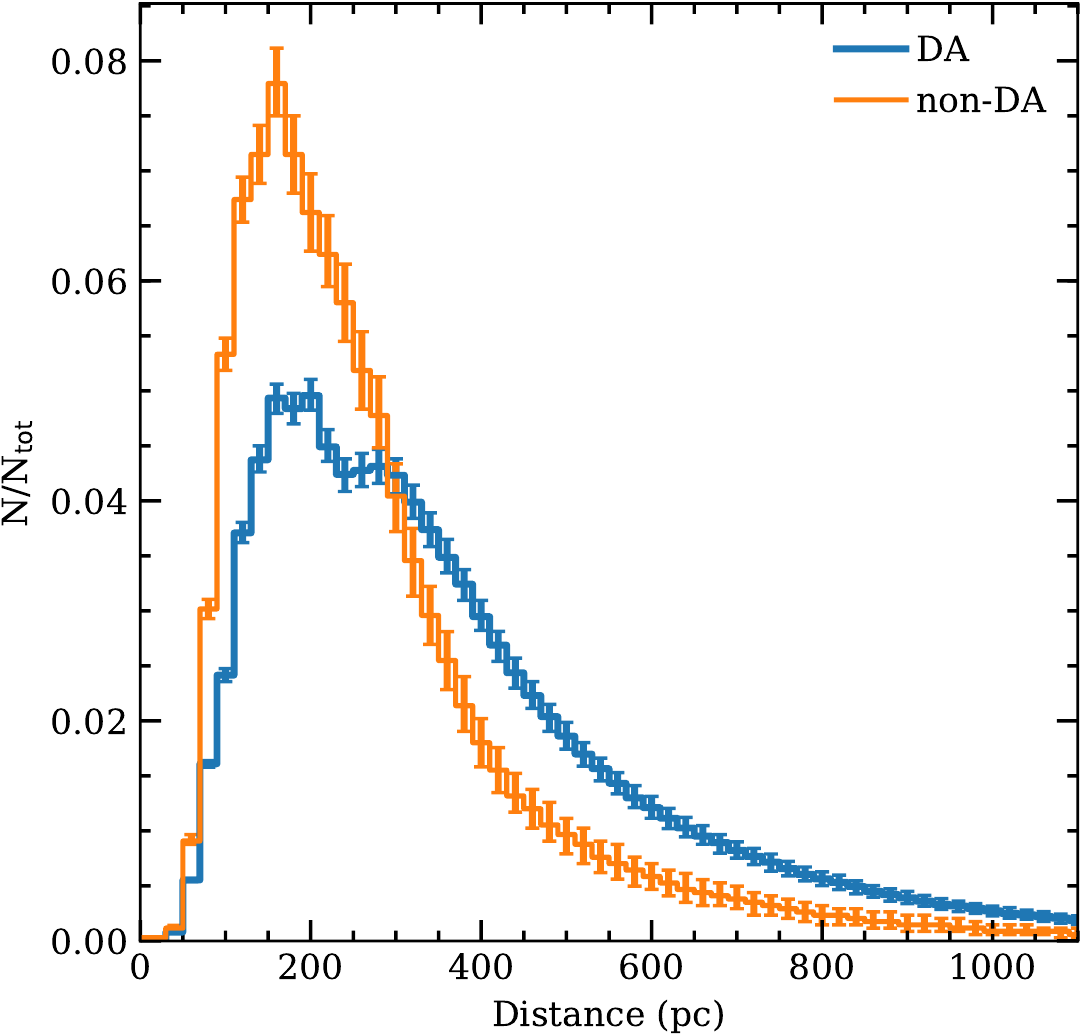}
	    	\caption{Distance distribution for DAs (blue) and non-DAs (orange). The non-DA distribution shows significantly more stars concentrated around $200\,\text{pc}$ than the DAs distribution, which shows an extended tail to higher distances.}
		\label{fig:DistanceDistribution}
	\end{figure}

	We compute the distance distribution of DAs and non-DAs using the distances determined from \textit{Gaia} parallaxes. As can be seen in Figure \ref{fig:DistanceDistribution}, the non-DA distribution is narrower than the DA distribution. The DA distribution has an extended tail to higher distances.
	As we are working with cool WDs that have spectroscopy in SDSS DR12, we expect a lower completeness for objects at higher distances or lower luminosities plus the saturation limit at magnitudes ${\sim}14.5$ on the $g$ filter.
	
	The scale-height of the galactic disk is around $300\,\text{pc}$~\citep{2017ASPC..509..421K} and the peaks of both of our distance distributions are around $200\,\text{pc}$. Thus we notice that our analysis could be influenced by the incompleteness of our sample. The cumulative distance distribution for our DA and non-DA samples follow a cubic rule up to $110\,\text{pc}$, indicating that our sample should be near complete up to this distance in the SDSS spectroscopic cone~\citep{2016arXiv160601236O, 2016MNRAS.462.2295H}. To ensure the reliability of our results we present our determinations for a close to complete sample corrected by the observed volume, following \citet{1968ApJ...151..393S, 1975ApJ...202...22S, 1980ApJ...238..685G, 1989MNRAS.238..709S, 2003AJ....125..348L, 2007MNRAS.375.1315K, 2010ApJ...714.1037L, 2015MNRAS.452.1637R}, limiting the magnitude on SDSS filter $g$, $14.5 < m_g < 19.0$, and assuming a galactic disk scale height of $300\,\text{pc}$. The resulting sample after applying the magnitude limits for the correction is composed of $5\,336$ DAs and $1\,315$ non-DAs.
	
	\subsection{Volume Corrected Distributions}
	
\begin{figure}
		\includegraphics[width=\columnwidth]{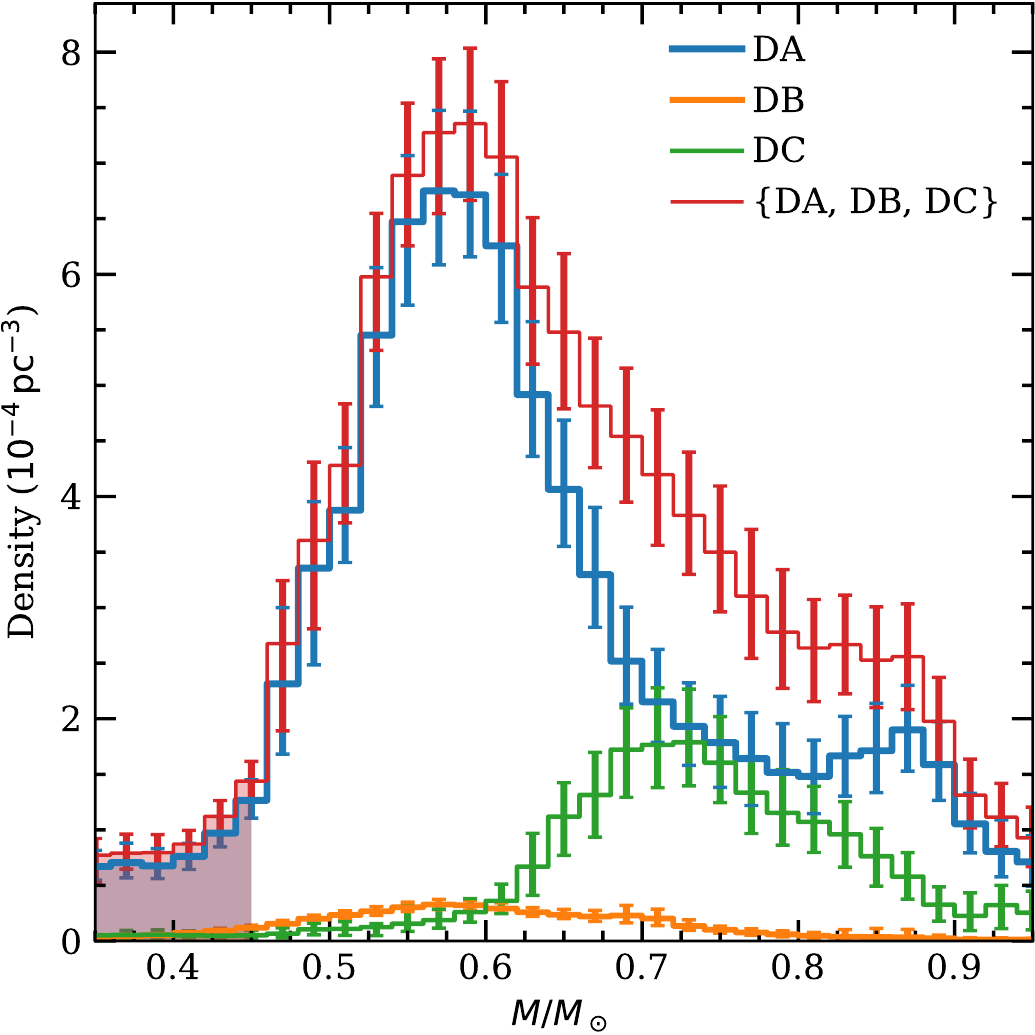}
	    	\caption{Mass distribution for DAs (blue), DBs (orange), DCs (green) and the sum of the three distributions (red). The peak at higher masses in the DA distribution indicates a population that could have come from mergers. Due their higher luminosities, DBs are not significant in volume corrected distribution. On the other hand, due the low radius and effective temperature of DCs, they correspond to $36$ per cent of the density of WDs above ${\sim}0.7\,\mathrm{M}_\odot$, but only $12$ per cent in number.}
		\label{fig:VmaxMass}
	\end{figure}

	Figure \ref{fig:VmaxMass} shows the volume corrected mass distribution for DAs, DBs and DCs, as well as the sum of the three distributions. In this figure we see an overdensity of DAs for masses higher than  ${\sim}0.75\,\mathrm{M}_\odot$. This overdensity could be explained if a burst of star formation occurred within the last $0.5\,\text{Gyr}$, considering single stellar evolution, since it shows a stellar mass distribution concentrated around ${\sim}0.87\,\mathrm{M}_\odot$. This scenario is not very probable, since the number of objects that should be formed in less than $0.5\,\text{Gyr}$ is too high. A more likely scenario involves mergers, that result in high mass WDs. The low mass binary component, also shown in Figure \ref{fig:distMass}, is not as prominent in Figure \ref{fig:VmaxMass}, since low mass WDs have higher radius and could be seen at much higher distances. Also we can see that DBs are not very significant for the total distribution, which is expected considering that they are intrinsically hot WDs. For DCs, we see an important contribution to the total number for lower effective temperatures, showing an increase in the number of objects for higher masses. Note that the number of DAs decreases in the same way as the number of DCs increases, indicating that some massive DAs could be turning into massive DCs.
	
\begin{figure}
		\includegraphics[width=\columnwidth]{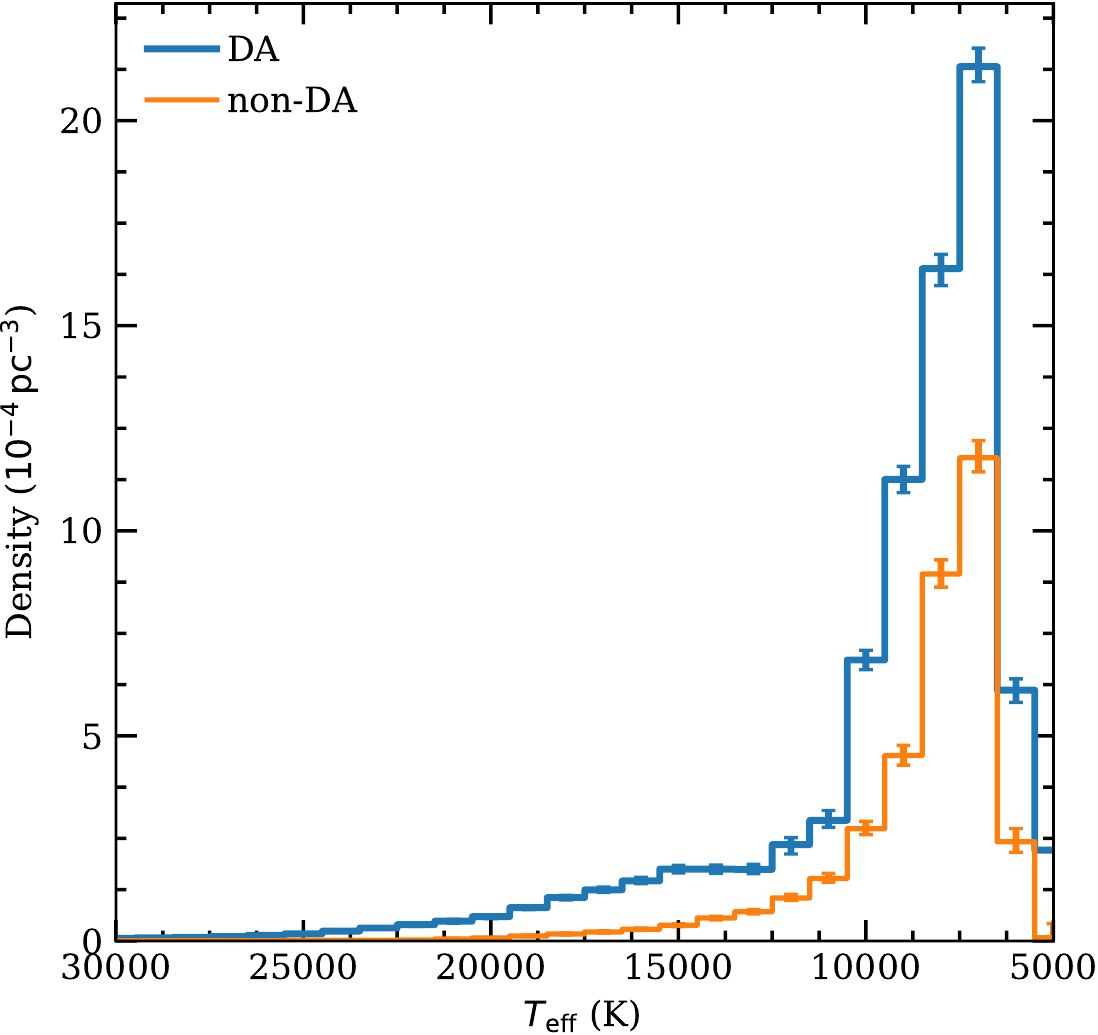}
	    	\caption{Photometric effective temperature distribution for DAs (blue) and non-DAs (orange). The decrease in the density of objects for both distributions around $5\,000\,\text{K}$ indicates that they have the same maximum age.}
		\label{fig:VmaxNDADB}
	\end{figure}

	Figure \ref{fig:VmaxNDADB} shows the volume corrected effective temperature distributions for DA and non-DAs. From this figure we see that both distributions show a decrease in the density around $5\,000\,\text{K}$, and present the same slope, indicating that the two samples have nearly the same age. Also we determine from the volume corrected effective temperature distribution that around $30{\pm}1$ per cent of the total WDs are non-DAs.
	
\begin{figure}
		\includegraphics[width=\columnwidth]{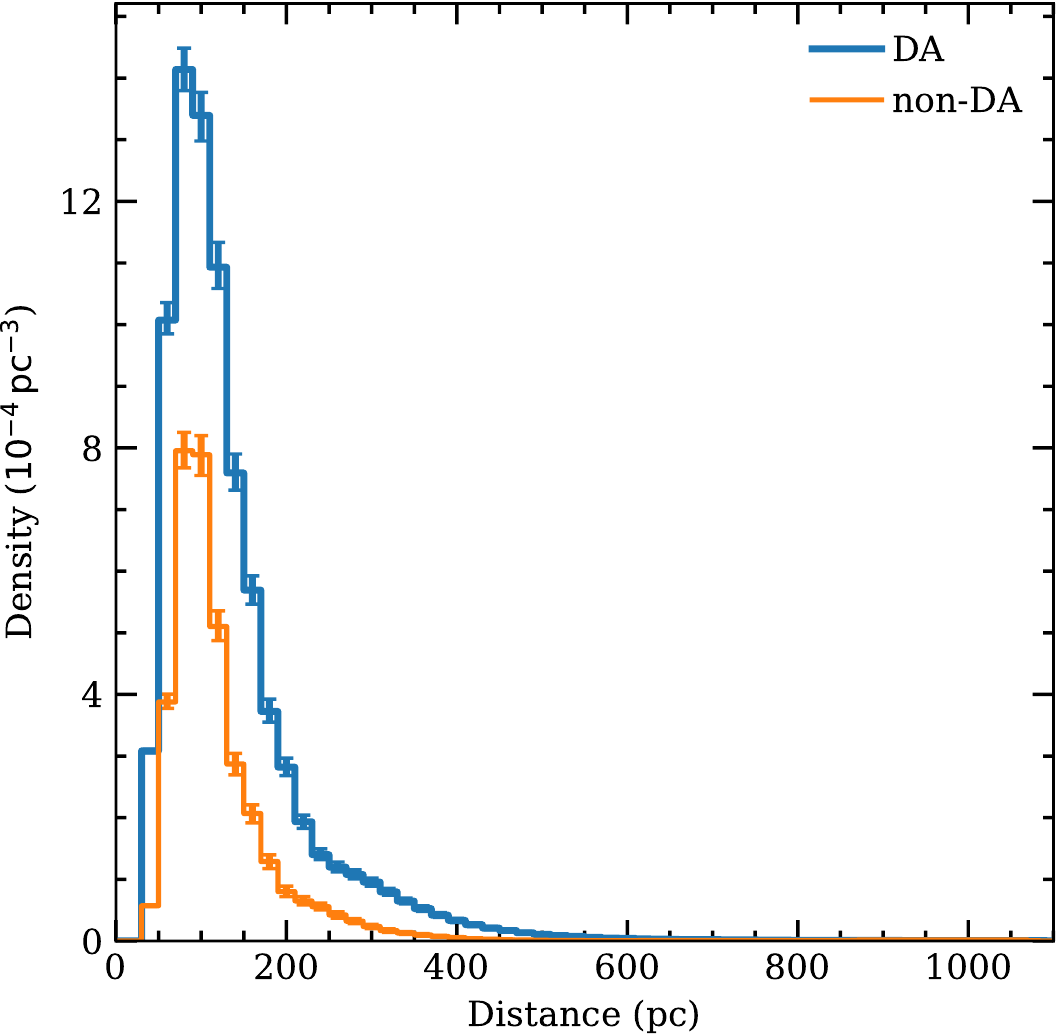}
	    	\caption{Distance distribution for DAs (blue) and non-DAs (orange) WDs after applying the volume correction. Our sample do not present any object closer than $30~\text{pc}$ due the SDSS saturation limits. Both distributions are normalised by the sum of weights resulting from the correction.}
		\label{fig:VmaxDistDist}
	\end{figure}

	In Figure \ref{fig:VmaxDistDist} we show the volume corrected distance distribution for DAs and non-DAs after applying the volume correction, normalised by the sum of weights resulting from the correction. We can see that DAs and non-DAs follow nearly the same distance distribution.
        
	For a uniform distribution along the galactic disk, we expect the mean of the volume calculated at each star distance over the maximum volume at which this star could be seen, i.e $V/V_{\mathrm{max}}$, to be around $0.5$. For our DAs and non-DAs samples we found that $V/V_{\mathrm{max}}$ is, respectively, $0.51$ and $0.50$. This means that within these magnitude limits both distributions follow a nearly uniform distribution.

\section{HELIUM TO HYDROGEN ATMOSPHERE WD NUMBER AND DENSITY RATIO}\label{sec:hehratio}

	An efficient way to study the difference between the DAs and non-DAs is through the helium to hydrogen number ratio, defined as the ratio between the number or density of helium-rich to hydrogen-rich atmosphere WDs as a function of the effective temperature. This quantity will give information on the differences on the formation and evolutionary channels of these two populations of WDs \citep{2008ApJ...672.1144T}.

\begin{figure}
		\includegraphics[width=\columnwidth]{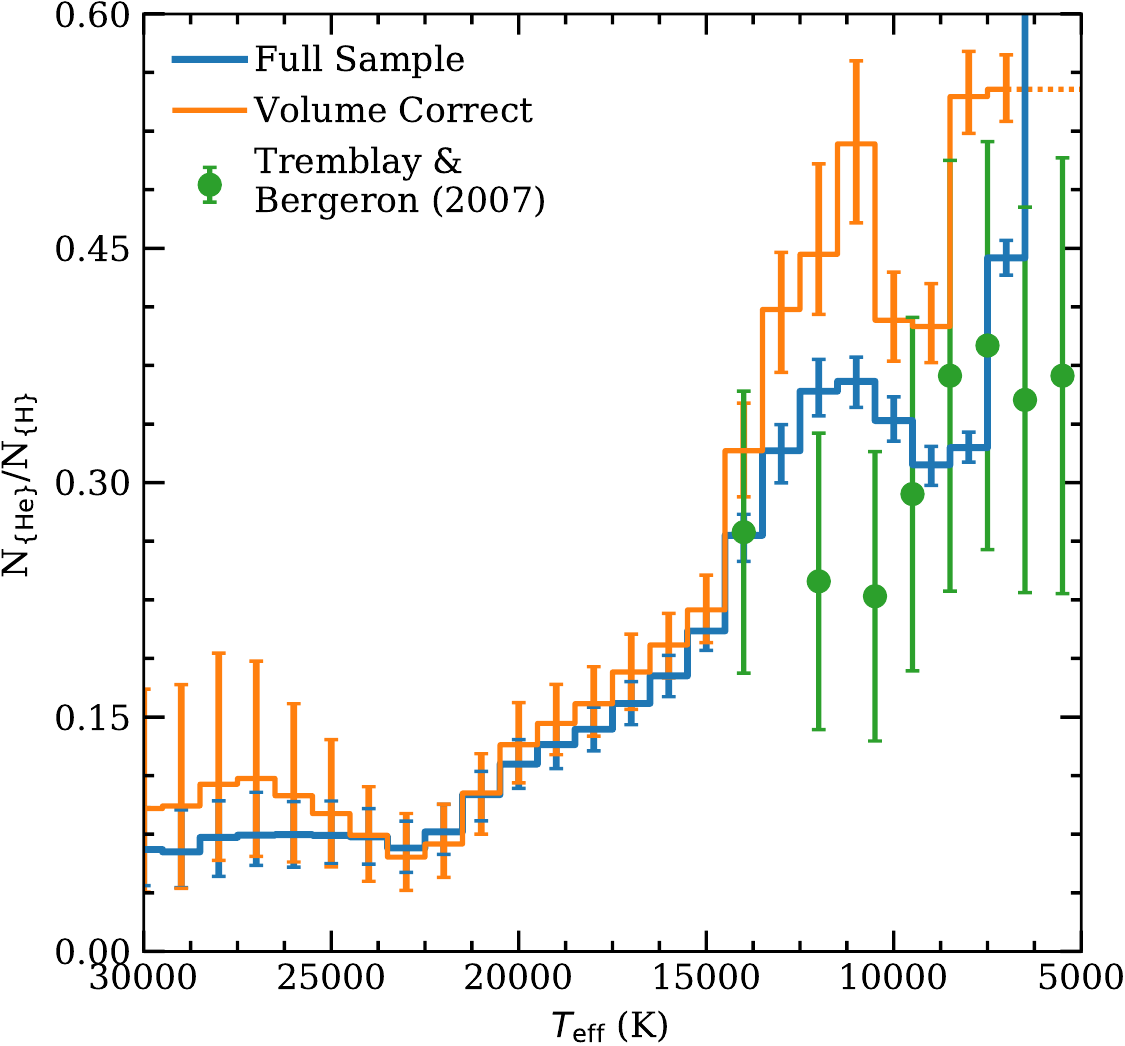}
	    	\caption{Helium to hydrogen atmosphere WD number ratio as a function of the effective temperature calculated with our final sample of $12\,811$ DAs and $3\,374$ non-DAs (blue), the density ratio calculated with our volume corrected sample of $4\,775$ DAs and $1308$ non-DAs (orange) and the ratio calculated by \citet{2008ApJ...672.1144T} (green). The effective temperature increases from right to left. Our determinations are for effective temperatures in the range  $30\, 000\,\text{K}{>}\mathrm{T}_{\rm{eff}}{>}5\,000\,\text{K}$, while the determination of \citet{2008ApJ...672.1144T} is in the range $15\, 000\,\text{K}{>}\mathrm{T}_{\rm{eff}}{>}5\,000\,\text{K}$. Our determinations show that the helium to hydrogen atmosphere WD number ratio shows an increase of more than $4$ times the ratio for effective temperatures between $21\,000\,\text{K}$ and $12\,500\,\text{K}$, followed by a decrease around ${\approx}10\,000\,\text{K}$ and then a large increase around $5\, 000\,\text{K}$, where DAs and non-DAs become practically indistinguishable.}
	\label{fig:HeHRatio}
	\end{figure}

	Figure \ref{fig:HeHRatio} depicts the helium to hydrogen atmosphere WD number ratio calculated with our final sample of $13\,687$ DAs and $3\,437$ non-DAs (blue line) and the density ratio calculated with our volume corrected sample of $5\,336$ DAs and $1\,315$ non-DAs (orange line).
	Also shown in Figure \ref{fig:HeHRatio} is the ratio calculated by \citet{2008ApJ...672.1144T} (green points).
	Note that, for both of our determinations, the helium to hydrogen atmosphere WD number ratio starts to increase significantly at effective temperatures around $22\, 000\,\text{K}$, where the convection in the helium layers becomes adiabatic~\citep[e.g.][]{2010A&ARv..18..471A}. Figure \ref{fig:HeHRatio} also presents a second increase around ${\approx}14\,000\,\text{K}$, which coincides with the temperature when the hydrogen convective layer starts to develop according to theoretical models. This increase is followed by a decrease around ${\approx}10\,000\,\text{K}$ and then a large increase around $5\, 000\,\text{K}$, where DAs and non-DAs become practically indistinguishable in optical spectra. \citet{2008ApJ...672.1144T} computed the helium to hydrogen atmosphere WD number ratio only in the effective temperature range $15\, 000\,\text{K}{>}\mathrm{T}_{\rm{eff}}{>}5\,000\,\text{K}$, using a sample of $340$ DAs and $107$ non-DAs (green dots in Figure \ref{fig:HeHRatio}), and found a factor of two increase in the helium to hydrogen atmosphere WD number ratio for effective temperatures around $10\, 000\,\text{K}$, which they explained by stating that $15$ per cent of the DAs turn into non-DAs for effective temperatures between $15\,000\,\text{K}$ and $10\,000\,\text{K}$. This result could be explained by the occurrence of convective mixing in the outer layers for WDs with hydrogen content, $\mathrm{M}_\mathrm{H}/\mathrm{M}_*$, below $10^{-8}$.
	
	Our determination of the helium to hydrogen atmosphere WD number and density ratio in the same effective temperatures range studied by \citet{2008ApJ...672.1144T} does not lead to the same conclusion, since the increase in the ratio around $10\,000\,\text{K}$ that they found is not seen in our data. We believe this effect could be related to the small number of objects considered in their analysis or their choice of their upper effective temperature limit and not to a property of the helium to hydrogen atmosphere WD number ratio itself.

 	\citet{2018ApJ...857...56R} propose that the convective dilution of a hydrogen layer with $\log \mathrm{M}_\mathrm{H}/\mathrm{M}_\odot {\sim}-14$ will occur for effective temperatures near $20\,000\,\text{K}$ and no more than $32\,000,\text{K}$, turning DAs into non-DAs. In Figure \ref{fig:HeHRatio} we can see an increase helium to hydrogen atmosphere WD number and density ratio at effective temperatures $22\,000\,\text{K}$, which is possible evidence of convective dilution. Spectroscopic catalogues reported approximately $250$ DABs around $24\,000\,\text{K}$~\citep{2016MNRAS.455.3413K,2015MNRAS.446.4078K,2006ApJS..167...40E,2004ApJ...607..426K}, very close to where we found an increase in the counts of non-DAs. For the total sample (blue line in Figure \ref{fig:HeHRatio}) the observed ratio at $12\,500\,\text{K}$ can be explained if $11{\pm}1$ per cent of DAs turn into non-DAs, while for our volume corrected sample (orange line in Figure \ref{fig:HeHRatio}) the value is $14{\pm}3$ per cent. Also, the envelope models employed \citet{2018ApJ...857...56R} show that helium-rich DAs could turn into DCs by convective mixing when their effective temperature goes below $12\,000,\text{K}$. In Figure \ref{fig:NDANDB} we can see a nearly constant number of DAs between $14\,000\,\text{K}$ and $12\,000\,\text{K}$, being a further possible evidence of objects transitioning from DAs to DBs due the convective mixing. It is also important to notice that spectroscopic WD catalogues report approximately $150$ DBAs, objects with spectra showing both hydrogen and helium lines, concentrated around $13\, 000\,\text{K}$~\citep{2016MNRAS.455.3413K,2015MNRAS.446.4078K,2006ApJS..167...40E,2004ApJ...607..426K}, being another possible evidence of this transition due to the convective mixing.

	Another hypothesis to explain the observed helium to hydrogen atmosphere WD number and density ratio behaviour is related to the different cooling times for different atmosphere compositions \citep{2001ApJS..133..413B}. Helium-rich atmosphere WDs are expected to cool faster since helium is less opaque than hydrogen. We calculated the ratio using the hydrogen-rich and helium-rich atmosphere WD fully evolutionary models from \citet{2015MNRAS.450.3708R} and found that the helium-rich evolutionary models would need to cool down at least three times faster than predicted until reaching $16\, 000\,\text{K}$ to reproduce the observed ratio, which is unlikely.
    
	Finally, we consider the possibility of multiple populations of WD stars, one of them with a higher probability of producing non-DAs. The nearly constant number of DAs between $14\,000\,\text{K}$ and $12\,000\,\text{K}$ could be explained by two populations, one only with objects older than $1\,\text{Gyr}$ and the second with most younger objects than $0.75\,\text{Gyr}$. This scenario is not very probable, however, it still could reproduce the observed ratio.
	
\section{CONCLUSIONS}\label{sec:conclusion}

	We find that DAs and DBs show nearly the same mass distribution, both having a peak around ${\sim}0.55\,\mathrm{M}_\odot$. Also, the mass distributions have nearly the same shape between ${\sim}0.45\,\mathrm{M}_\odot$ and ${\sim}0.95\,\mathrm{M}_\odot$. However, the mass distribution for DAs shows an  excess in the number of objects at higher and lower masses. This could be an indication of a population of DA WDs originating from interacting binaries. Also, we found that the DCs mass distribution has a peak at ${\sim}0.7\,\mathrm{M}_\odot$, which is ${\sim}0.15\,\mathrm{M}_\odot$ higher that the one observed in the DAs and DBs distributions.
	
	Our calculations of the helium to hydrogen number ratio shows an increase from ${\sim}0.075$ to $\sim{0.36}$ at effective temperatures from $22\,000\,\text{K}$ to $12\,000\,\text{K}$. The main hypothesis to explain this increase are: convective dilution and mixing, cooling rates with high dependence of the atmosphere composition and multiple population scenario.
	
	Contrary to the results of \citet{2008ApJ...672.1144T}, we do not find an increase of the helium to hydrogen number ratio for effective temperatures below $10\,000\,\text{K}$. This is related to the fact that the sample considered in this work is significantly larger than that used by \citet{2008ApJ...672.1144T}. Also, we extend the effective temperature range up to $30\, 000$ K.
                
        Our study indicates that the differences in the cooling for hydrogen-rich and helium-rich atmosphere WDs evolutionary models are not enough to explain the observed ratio. Also, to reproduce the ratio with multiple populations, we need at least two distinct population, where one of them has the total age smaller than $0.75\,\text{Gyr}$.

        The most acceptable hypothesis to explain the increase of the helium to hydrogen atmosphere WD number ratio between $21\,000\,\text{K}$ and $15\,000\,\text{K}$ is the convective dilution of the hydrogen layers turning around $14{\pm}3$ per cent of DAs into non-DAs. The increase from $15\,000\,\text{K}$ to $12\,000\,\text{K}$ is a consequence of the convective mixing\citep[see][]{2018ApJ...857...56R}. Also the estimated mass distribution for DCs indicates that non-DAs cooler than $10\,000\,\text{K}$ are on average ${\sim}0.15\,\mathrm{M}_\odot$ more massive than DAs in the same effective temperature range, which is expected since the more massive DAs have thin hydrogen layers, making the convective mixing and dilution more likely to occur~\citep{2012MNRAS.420.1462R}.




\section*{Acknowledgements}

We thank the referee for his suggestions and comments, that improved the paper.

GO, ADR, SOK and LAA received support from from CNPq and PRONEX-FAPERGS/CNPq (Brazil). DK received
support from programme  Science  without  Borders,  MCIT/MEC-Brazil. 
This research has made use of NASA Astrophysics Data System. 

Funding for SDSS-III has been provided by the Alfred P. Sloan Foundation, the Participating Institutions,
the National Science Foundation, and the U.S. Department of Energy Office of Science. The SDSS-III web site
is http://www.sdss3.org/. 

SDSS-III is managed by the Astrophysical Research
Consortium for the Participating Institutions of the SDSS-III Collaboration including the University of Arizona, the
Brazilian Participation Group, Brookhaven National Laboratory, Carnegie Mellon University, University of Florida,
the French Participation Group, the German Participation
Group, Harvard University, the Instituto de Astrofísica de
Canarias, the Michigan State/Notre Dame/JINA Participation Group, Johns Hopkins University, Lawrence 
Berkeley National Laboratory, Max Planck Institute for Astrophysics, Max Planck Institute for Extraterrestrial Physics,
New Mexico State University, New York University, Ohio State University, Pennsylvania State University, University
of Portsmouth, Princeton University, the Spanish Participation Group, University of Tokyo, University of Utah,
Vanderbilt University, University of Virginia, University of Washington, and Yale University.

This work has made use of data from the European Space Agency (ESA)
mission {\it Gaia} (\url{https://www.cosmos.esa.int/gaia}), processed by
the {\it Gaia} Data Processing and Analysis Consortium (DPAC,
\url{https://www.cosmos.esa.int/web/gaia/dpac/consortium}). Funding
for the DPAC has been provided by national institutions, in particular
the institutions participating in the {\it Gaia} Multilateral Agreement.

GO thanks Anna and her cat, Cosmos, for the company and support on this work.

\bibliographystyle{mnras}

\begin{thebibliography}{}
\makeatletter
\relax
\def\mn@urlcharsother{\let\do\@makeother \do\$\do\&\do\#\do\^\do\_\do\%\do\~}
\def\mn@doi{\begingroup\mn@urlcharsother \@ifnextchar [ {\mn@doi@}
  {\mn@doi@[]}}
\def\mn@doi@[#1]#2{\def\@tempa{#1}\ifx\@tempa\@empty \href
  {http://dx.doi.org/#2} {doi:#2}\else \href {http://dx.doi.org/#2} {#1}\fi
  \endgroup}
\def\mn@eprint#1#2{\mn@eprint@#1:#2::\@nil}
\def\mn@eprint@arXiv#1{\href {http://arxiv.org/abs/#1} {{\tt arXiv:#1}}}
\def\mn@eprint@dblp#1{\href {http://dblp.uni-trier.de/rec/bibtex/#1.xml}
  {dblp:#1}}
\def\mn@eprint@#1:#2:#3:#4\@nil{\def\@tempa {#1}\def\@tempb {#2}\def\@tempc
  {#3}\ifx \@tempc \@empty \let \@tempc \@tempb \let \@tempb \@tempa \fi \ifx
  \@tempb \@empty \def\@tempb {arXiv}\fi \@ifundefined
  {mn@eprint@\@tempb}{\@tempb:\@tempc}{\expandafter \expandafter \csname
  mn@eprint@\@tempb\endcsname \expandafter{\@tempc}}}

\bibitem[\protect\citeauthoryear{{Abolfathi} et~al.,}{{Abolfathi}
  et~al.}{2017}]{2017arXiv170709322A}
{Abolfathi} B.,  et~al., 2017, preprint, \href
  {http://adsabs.harvard.edu/abs/2017arXiv170709322A} {} (\mn@eprint {arXiv}
  {1707.09322})

\bibitem[\protect\citeauthoryear{{Althaus}, {C{\'o}rsico}, {Isern}  \&
  {Garc{\'{\i}}a-Berro}}{{Althaus} et~al.}{2010}]{2010A&ARv..18..471A}
{Althaus} L.~G.,  {C{\'o}rsico} A.~H.,  {Isern} J.,   {Garc{\'{\i}}a-Berro} E.,
   2010, \mn@doi [\aapr] {10.1007/s00159-010-0033-1}, \href
  {http://adsabs.harvard.edu/abs/2010A%26ARv..18..471A} {18, 471}

\bibitem[\protect\citeauthoryear{{Bergeron}, {Ruiz}  \& {Leggett}}{{Bergeron}
  et~al.}{1997}]{1997ApJS..108..339B}
{Bergeron} P.,  {Ruiz} M.~T.,   {Leggett} S.~K.,  1997, \mn@doi [\apjs]
  {10.1086/312955}, \href {http://adsabs.harvard.edu/abs/1997ApJS..108..339B}
  {108, 339}

\bibitem[\protect\citeauthoryear{{Bergeron}, {Leggett}  \& {Ruiz}}{{Bergeron}
  et~al.}{2001}]{2001ApJS..133..413B}
{Bergeron} P.,  {Leggett} S.~K.,   {Ruiz} M.~T.,  2001, \mn@doi [\apjs]
  {10.1086/320356}, \href {http://adsabs.harvard.edu/abs/2001ApJS..133..413B}
  {133, 413}

\bibitem[\protect\citeauthoryear{{Bono}, {Salaris}  \& {Gilmozzi}}{{Bono}
  et~al.}{2013}]{2013A&A...549A.102B}
{Bono} G.,  {Salaris} M.,   {Gilmozzi} R.,  2013, \mn@doi [\aap]
  {10.1051/0004-6361/201220024}, \href
  {http://adsabs.harvard.edu/abs/2013A%26A...549A.102B} {549, A102}

\bibitem[\protect\citeauthoryear{{Brassard}, {Fontaine}, {Dufour}  \&
  {Bergeron}}{{Brassard} et~al.}{2007}]{2007ASPC..372...19B}
{Brassard} P.,  {Fontaine} G.,  {Dufour} P.,   {Bergeron} P.,  2007, in
  {Napiwotzki} R.,  {Burleigh} M.~R.,  eds,  Astronomical Society of the
  Pacific Conference Series Vol. 372, 15th European Workshop on White Dwarfs.
  p.~19

\bibitem[\protect\citeauthoryear{{Campos} et~al.,}{{Campos}
  et~al.}{2016}]{2016MNRAS.456.3729C}
{Campos} F.,  et~al., 2016, \mn@doi [\mnras] {10.1093/mnras/stv2911}, \href
  {http://adsabs.harvard.edu/abs/2016MNRAS.456.3729C} {456, 3729}

\bibitem[\protect\citeauthoryear{{Carrasco}, {Catal{\'a}n}, {Jordi},
  {Tremblay}, {Napiwotzki}, {Luri}, {Robin}  \& {Kowalski}}{{Carrasco}
  et~al.}{2014}]{2014A&A...565A..11C}
{Carrasco} J.~M.,  {Catal{\'a}n} S.,  {Jordi} C.,  {Tremblay} P.-E.,
  {Napiwotzki} R.,  {Luri} X.,  {Robin} A.~C.,   {Kowalski} P.~M.,  2014,
  \mn@doi [\aap] {10.1051/0004-6361/201220596}, \href
  {http://adsabs.harvard.edu/abs/2014A%26A...565A..11C} {565, A11}

\bibitem[\protect\citeauthoryear{{Castanheira} \& {Kepler}}{{Castanheira} \&
  {Kepler}}{2009}]{2009AIPC.1170..616C}
{Castanheira} B.~G.,  {Kepler} S.~O.,  2009, in {Guzik} J.~A.,  {Bradley}
  P.~A.,  eds,  American Institute of Physics Conference Series Vol. 1170,
  American Institute of Physics Conference Series. pp 616--620,
  \mn@doi{10.1063/1.3246572}

\bibitem[\protect\citeauthoryear{{Doherty}, {Gil-Pons}, {Siess}, {Lattanzio}
  \& {Lau}}{{Doherty} et~al.}{2015}]{2015MNRAS.446.2599D}
{Doherty} C.~L.,  {Gil-Pons} P.,  {Siess} L.,  {Lattanzio} J.~C.,   {Lau}
  H.~H.~B.,  2015, \mn@doi [\mnras] {10.1093/mnras/stu2180}, \href
  {http://adsabs.harvard.edu/abs/2015MNRAS.446.2599D} {446, 2599}

\bibitem[\protect\citeauthoryear{{Doi} et~al.,}{{Doi}
  et~al.}{2010}]{2010AJ....139.1628D}
{Doi} M.,  et~al., 2010, \mn@doi [\aj] {10.1088/0004-6256/139/4/1628}, \href
  {http://adsabs.harvard.edu/abs/2010AJ....139.1628D} {139, 1628}

\bibitem[\protect\citeauthoryear{{Dufour}, {Bergeron}  \& {Fontaine}}{{Dufour}
  et~al.}{2005}]{2005ASPC..334..197D}
{Dufour} P.,  {Bergeron} P.,   {Fontaine} G.,  2005, in {Koester} D.,
  {Moehler} S.,  eds,  Astronomical Society of the Pacific Conference Series
  Vol. 334, 14th European Workshop on White Dwarfs. p.~197 (\mn@eprint {}
  {astro-ph/0503448})

\bibitem[\protect\citeauthoryear{{Eisenstein} et~al.,}{{Eisenstein}
  et~al.}{2006}]{2006ApJS..167...40E}
{Eisenstein} D.~J.,  et~al., 2006, \mn@doi [\apjs] {10.1086/507110}, \href
  {http://adsabs.harvard.edu/abs/2006ApJS..167...40E} {167, 40}

\bibitem[\protect\citeauthoryear{{Fontaine} \& {Brassard}}{{Fontaine} \&
  {Brassard}}{2008}]{2008PASP..120.1043F}
{Fontaine} G.,  {Brassard} P.,  2008, \mn@doi [\pasp] {10.1086/592788}, \href
  {http://adsabs.harvard.edu/abs/2008PASP..120.1043F} {120, 1043}

\bibitem[\protect\citeauthoryear{{G{\"a}nsicke}, {Koester}, {Girven}, {Marsh}
  \& {Steeghs}}{{G{\"a}nsicke} et~al.}{2010}]{2010Sci...327..188G}
{G{\"a}nsicke} B.~T.,  {Koester} D.,  {Girven} J.,  {Marsh} T.~R.,   {Steeghs}
  D.,  2010, \mn@doi [Science] {10.1126/science.1180228}, \href
  {http://adsabs.harvard.edu/abs/2010Sci...327..188G} {327, 188}

\bibitem[\protect\citeauthoryear{{Garcia-Berro}, {Hernanz}, {Mochkovitch}  \&
  {Isern}}{{Garcia-Berro} et~al.}{1988a}]{1988A&A...193..141G}
{Garcia-Berro} E.,  {Hernanz} M.,  {Mochkovitch} R.,   {Isern} J.,  1988a,
  \aap, \href {http://adsabs.harvard.edu/abs/1988A%26A...193..141G} {193, 141}

\bibitem[\protect\citeauthoryear{{Garcia-Berro}, {Hernanz}, {Isern}  \&
  {Mochkovitch}}{{Garcia-Berro} et~al.}{1988b}]{1988Natur.333..642G}
{Garcia-Berro} E.,  {Hernanz} M.,  {Isern} J.,   {Mochkovitch} R.,  1988b,
  \mn@doi [\nat] {10.1038/333642a0}, \href
  {http://adsabs.harvard.edu/abs/1988Natur.333..642G} {333, 642}

\bibitem[\protect\citeauthoryear{{Gentile Fusillo}, {G{\"a}nsicke}, {Farihi},
  {Koester}, {Schreiber}  \& {Pala}}{{Gentile Fusillo}
  et~al.}{2017}]{2017MNRAS.468..971G}
{Gentile Fusillo} N.~P.,  {G{\"a}nsicke} B.~T.,  {Farihi} J.,  {Koester} D.,
  {Schreiber} M.~R.,   {Pala} A.~F.,  2017, \mn@doi [\mnras]
  {10.1093/mnras/stx468}, \href
  {http://adsabs.harvard.edu/abs/2017MNRAS.468..971G} {468, 971}

\bibitem[\protect\citeauthoryear{{Green}}{{Green}}{1980}]{1980ApJ...238..685G}
{Green} R.~F.,  1980, \mn@doi [\apj] {10.1086/158025}, \href
  {http://adsabs.harvard.edu/abs/1980ApJ...238..685G} {238, 685}

\bibitem[\protect\citeauthoryear{{Gunn} et~al.,}{{Gunn}
  et~al.}{1998}]{1998AJ....116.3040G}
{Gunn} J.~E.,  et~al., 1998, \mn@doi [\aj] {10.1086/300645}, \href
  {http://adsabs.harvard.edu/abs/1998AJ....116.3040G} {116, 3040}

\bibitem[\protect\citeauthoryear{{Hamada} \& {Salpeter}}{{Hamada} \&
  {Salpeter}}{1961}]{1961ApJ...134..683H}
{Hamada} T.,  {Salpeter} E.~E.,  1961, \mn@doi [\apj] {10.1086/147195}, \href
  {http://adsabs.harvard.edu/abs/1961ApJ...134..683H} {134, 683}

\bibitem[\protect\citeauthoryear{{Hansen} et~al.,}{{Hansen}
  et~al.}{2002}]{2002ApJ...574L.155H}
{Hansen} B.~M.~S.,  et~al., 2002, \mn@doi [\apjl] {10.1086/342528}, \href
  {http://adsabs.harvard.edu/abs/2002ApJ...574L.155H} {574, L155}

\bibitem[\protect\citeauthoryear{{Hansen} et~al.,}{{Hansen}
  et~al.}{2007}]{2007ApJ...671..380H}
{Hansen} B.~M.~S.,  et~al., 2007, \mn@doi [\apj] {10.1086/522567}, \href
  {http://adsabs.harvard.edu/abs/2007ApJ...671..380H} {671, 380}

\bibitem[\protect\citeauthoryear{{Holberg}, {Oswalt}, {Sion}  \&
  {McCook}}{{Holberg} et~al.}{2016}]{2016MNRAS.462.2295H}
{Holberg} J.~B.,  {Oswalt} T.~D.,  {Sion} E.~M.,   {McCook} G.~P.,  2016,
  \mn@doi [\mnras] {10.1093/mnras/stw1357}, \href
  {http://adsabs.harvard.edu/abs/2016MNRAS.462.2295H} {462, 2295}

\bibitem[\protect\citeauthoryear{{Ibeling} \& {Heger}}{{Ibeling} \&
  {Heger}}{2013}]{2013ApJ...765L..43I}
{Ibeling} D.,  {Heger} A.,  2013, \mn@doi [\apjl]
  {10.1088/2041-8205/765/2/L43}, \href
  {http://adsabs.harvard.edu/abs/2013ApJ...765L..43I} {765, L43}

\bibitem[\protect\citeauthoryear{{Isern}, {Garc{\'{\i}}a-Berro}  \&
  {Salaris}}{{Isern} et~al.}{2001}]{2001ASPC..245..328I}
{Isern} J.,  {Garc{\'{\i}}a-Berro} E.,   {Salaris} M.,  2001, in {von Hippel}
  T.,  {Simpson} C.,   {Manset} N.,  eds,  Astronomical Society of the Pacific
  Conference Series Vol. 245, Astrophysical Ages and Times Scales. p.~328

\bibitem[\protect\citeauthoryear{{Kalirai}, {Ventura}, {Richer}, {Fahlman},
  {Durrell}, {D'Antona}  \& {Marconi}}{{Kalirai}
  et~al.}{2001}]{2001AJ....122.3239K}
{Kalirai} J.~S.,  {Ventura} P.,  {Richer} H.~B.,  {Fahlman} G.~G.,  {Durrell}
  P.~R.,  {D'Antona} F.,   {Marconi} G.,  2001, \mn@doi [\aj] {10.1086/324463},
  \href {http://adsabs.harvard.edu/abs/2001AJ....122.3239K} {122, 3239}

\bibitem[\protect\citeauthoryear{{Kalirai} et~al.,}{{Kalirai}
  et~al.}{2013}]{2013ApJ...763..110K}
{Kalirai} J.~S.,  et~al., 2013, \mn@doi [\apj] {10.1088/0004-637X/763/2/110},
  \href {http://adsabs.harvard.edu/abs/2013ApJ...763..110K} {763, 110}

\bibitem[\protect\citeauthoryear{{Kepler}, {Kleinman}, {Nitta}, {Koester},
  {Castanheira}, {Giovannini}, {Costa}  \& {Althaus}}{{Kepler}
  et~al.}{2007}]{2007MNRAS.375.1315K}
{Kepler} S.~O.,  {Kleinman} S.~J.,  {Nitta} A.,  {Koester} D.,  {Castanheira}
  B.~G.,  {Giovannini} O.,  {Costa} A.~F.~M.,   {Althaus} L.,  2007, \mn@doi
  [\mnras] {10.1111/j.1365-2966.2006.11388.x}, \href
  {http://adsabs.harvard.edu/abs/2007MNRAS.375.1315K} {375, 1315}

\bibitem[\protect\citeauthoryear{{Kepler} et~al.,}{{Kepler}
  et~al.}{2015}]{2015MNRAS.446.4078K}
{Kepler} S.~O.,  et~al., 2015, \mn@doi [\mnras] {10.1093/mnras/stu2388}, \href
  {http://adsabs.harvard.edu/abs/2015MNRAS.446.4078K} {446, 4078}

\bibitem[\protect\citeauthoryear{{Kepler}, {Koester}  \& {Ourique}}{{Kepler}
  et~al.}{2016a}]{2016Sci...352...67K}
{Kepler} S.~O.,  {Koester} D.,   {Ourique} G.,  2016a, \mn@doi [Science]
  {10.1126/science.aad6705}, \href
  {http://adsabs.harvard.edu/abs/2016Sci...352...67K} {352, 67}

\bibitem[\protect\citeauthoryear{{Kepler} et~al.,}{{Kepler}
  et~al.}{2016b}]{2016MNRAS.455.3413K}
{Kepler} S.~O.,  et~al., 2016b, \mn@doi [\mnras] {10.1093/mnras/stv2526}, \href
  {http://adsabs.harvard.edu/abs/2016MNRAS.455.3413K} {455, 3413}

\bibitem[\protect\citeauthoryear{{Kepler}, {Koester}, {Romero}, {Ourique}  \&
  {Pelisoli}}{{Kepler} et~al.}{2017}]{2017ASPC..509..421K}
{Kepler} S.~O.,  {Koester} D.,  {Romero} A.~D.,  {Ourique} G.,   {Pelisoli} I.,
   2017, in {Tremblay} P.-E.,  {Gaensicke} B.,   {Marsh} T.,  eds,
  Astronomical Society of the Pacific Conference Series Vol. 509, 20th European
  White Dwarf Workshop. p.~421 (\mn@eprint {arXiv} {1610.00371})

\bibitem[\protect\citeauthoryear{{Kleinman} et~al.,}{{Kleinman}
  et~al.}{2004}]{2004ApJ...607..426K}
{Kleinman} S.~J.,  et~al., 2004, \mn@doi [\apj] {10.1086/383464}, \href
  {http://adsabs.harvard.edu/abs/2004ApJ...607..426K} {607, 426}

\bibitem[\protect\citeauthoryear{{Kleinman} et~al.,}{{Kleinman}
  et~al.}{2013}]{2013ApJS..204....5K}
{Kleinman} S.~J.,  et~al., 2013, \mn@doi [\apjs] {10.1088/0067-0049/204/1/5},
  \href {http://adsabs.harvard.edu/abs/2013ApJS..204....5K} {204, 5}

\bibitem[\protect\citeauthoryear{{Koester}}{{Koester}}{1978}]{1978A&A....64..289K}
{Koester} D.,  1978, \aap, \href
  {http://adsabs.harvard.edu/abs/1978A%26A....64..289K} {64, 289}

\bibitem[\protect\citeauthoryear{{Koester}}{{Koester}}{2010}]{2010MmSAI..81..921K}
{Koester} D.,  2010, \memsai, \href
  {http://adsabs.harvard.edu/abs/2010MmSAI..81..921K} {81, 921}

\bibitem[\protect\citeauthoryear{{Koester} \& {Kepler}}{{Koester} \&
  {Kepler}}{2015}]{2015A&A...583A..86K}
{Koester} D.,  {Kepler} S.~O.,  2015, \mn@doi [\aap]
  {10.1051/0004-6361/201527169}, \href
  {http://adsabs.harvard.edu/abs/2015A%26A...583A..86K} {583, A86}

\bibitem[\protect\citeauthoryear{{Koester}, {Weidemann}  \&
  {Zeidler}}{{Koester} et~al.}{1982}]{1982A&A...116..147K}
{Koester} D.,  {Weidemann} V.,   {Zeidler} E.-M.,  1982, \aap, \href
  {http://adsabs.harvard.edu/abs/1982A%26A...116..147K} {116, 147}

\bibitem[\protect\citeauthoryear{{Koester}, {Girven}, {G{\"a}nsicke}  \&
  {Dufour}}{{Koester} et~al.}{2011}]{2011A&A...530A.114K}
{Koester} D.,  {Girven} J.,  {G{\"a}nsicke} B.~T.,   {Dufour} P.,  2011,
  \mn@doi [\aap] {10.1051/0004-6361/201116816}, \href
  {http://adsabs.harvard.edu/abs/2011A%26A...530A.114K} {530, A114}

\bibitem[\protect\citeauthoryear{{Liebert}, {Bergeron}  \& {Holberg}}{{Liebert}
  et~al.}{2003}]{2003AJ....125..348L}
{Liebert} J.,  {Bergeron} P.,   {Holberg} J.~B.,  2003, \mn@doi [\aj]
  {10.1086/345573}, \href {http://adsabs.harvard.edu/abs/2003AJ....125..348L}
  {125, 348}

\bibitem[\protect\citeauthoryear{{Liebert}, {Bergeron}  \& {Holberg}}{{Liebert}
  et~al.}{2005}]{2005ApJS..156...47L}
{Liebert} J.,  {Bergeron} P.,   {Holberg} J.~B.,  2005, \mn@doi [\apjs]
  {10.1086/425738}, \href {http://adsabs.harvard.edu/abs/2005ApJS..156...47L}
  {156, 47}

\bibitem[\protect\citeauthoryear{{Limoges} \& {Bergeron}}{{Limoges} \&
  {Bergeron}}{2010}]{2010ApJ...714.1037L}
{Limoges} M.-M.,  {Bergeron} P.,  2010, \mn@doi [\apj]
  {10.1088/0004-637X/714/2/1037}, \href
  {http://adsabs.harvard.edu/abs/2010ApJ...714.1037L} {714, 1037}

\bibitem[\protect\citeauthoryear{{Luri} et~al.,}{{Luri}
  et~al.}{2018}]{2018arXiv180409376L}
{Luri} X.,  et~al., 2018, preprint, \href
  {http://adsabs.harvard.edu/abs/2018arXiv180409376L} {} (\mn@eprint {arXiv}
  {1804.09376})

\bibitem[\protect\citeauthoryear{{MacDonald}, {Hernanz}  \& {Jose}}{{MacDonald}
  et~al.}{1998}]{1998MNRAS.296..523M}
{MacDonald} J.,  {Hernanz} M.,   {Jose} J.,  1998, \mn@doi [\mnras]
  {10.1046/j.1365-8711.1998.01392.x}, \href
  {http://adsabs.harvard.edu/abs/1998MNRAS.296..523M} {296, 523}

\bibitem[\protect\citeauthoryear{{Munn} et~al.,}{{Munn}
  et~al.}{2014}]{2014AJ....148..132M}
{Munn} J.~A.,  et~al., 2014, \mn@doi [\aj] {10.1088/0004-6256/148/6/132}, \href
  {http://adsabs.harvard.edu/abs/2014AJ....148..132M} {148, 132}

\bibitem[\protect\citeauthoryear{{Munn} et~al.,}{{Munn}
  et~al.}{2017}]{2017AJ....153...10M}
{Munn} J.~A.,  et~al., 2017, \mn@doi [\aj] {10.3847/1538-3881/153/1/10}, \href
  {http://adsabs.harvard.edu/abs/2017AJ....153...10M} {153, 10}

\bibitem[\protect\citeauthoryear{{Oswalt}, {Sion}  \& {McCook}}{{Oswalt}
  et~al.}{2016}]{2016arXiv160601236O}
{Oswalt} J.~B.~H.~T.~D.,  {Sion} E.~M.,   {McCook} G.~P.,  2016, preprint,
  \href {http://adsabs.harvard.edu/abs/2016arXiv160601236O} {} (\mn@eprint
  {arXiv} {1606.01236})

\bibitem[\protect\citeauthoryear{{Paczy{\'n}ski}}{{Paczy{\'n}ski}}{1971}]{1971AcA....21..417P}
{Paczy{\'n}ski} B.,  1971, \actaa, \href
  {http://adsabs.harvard.edu/abs/1971AcA....21..417P} {21, 417}

\bibitem[\protect\citeauthoryear{{Pelletier}, {Fontaine}, {Wesemael}, {Michaud}
   \& {Wegner}}{{Pelletier} et~al.}{1986}]{1986ApJ...307..242P}
{Pelletier} C.,  {Fontaine} G.,  {Wesemael} F.,  {Michaud} G.,   {Wegner} G.,
  1986, \mn@doi [\apj] {10.1086/164410}, \href
  {http://adsabs.harvard.edu/abs/1986ApJ...307..242P} {307, 242}

\bibitem[\protect\citeauthoryear{{Provencal}, {Shipman}, {H{\o}g}  \&
  {Thejll}}{{Provencal} et~al.}{1998}]{1998ApJ...494..759P}
{Provencal} J.~L.,  {Shipman} H.~L.,  {H{\o}g} E.,   {Thejll} P.,  1998,
  \mn@doi [\apj] {10.1086/305238}, \href
  {http://adsabs.harvard.edu/abs/1998ApJ...494..759P} {494, 759}

\bibitem[\protect\citeauthoryear{{Rebassa-Mansergas}, {Nebot
  G{\'o}mez-Mor{\'a}n}, {Schreiber}, {Girven}  \&
  {G{\"a}nsicke}}{{Rebassa-Mansergas} et~al.}{2011}]{2011MNRAS.413.1121R}
{Rebassa-Mansergas} A.,  {Nebot G{\'o}mez-Mor{\'a}n} A.,  {Schreiber} M.~R.,
  {Girven} J.,   {G{\"a}nsicke} B.~T.,  2011, \mn@doi [\mnras]
  {10.1111/j.1365-2966.2011.18200.x}, \href
  {http://adsabs.harvard.edu/abs/2011MNRAS.413.1121R} {413, 1121}

\bibitem[\protect\citeauthoryear{{Rebassa-Mansergas}, {Rybicka}, {Liu}, {Han}
  \& {Garc{\'{\i}}a-Berro}}{{Rebassa-Mansergas}
  et~al.}{2015}]{2015MNRAS.452.1637R}
{Rebassa-Mansergas} A.,  {Rybicka} M.,  {Liu} X.-W.,  {Han} Z.,
  {Garc{\'{\i}}a-Berro} E.,  2015, \mn@doi [\mnras] {10.1093/mnras/stv1399},
  \href {http://adsabs.harvard.edu/abs/2015MNRAS.452.1637R} {452, 1637}

\bibitem[\protect\citeauthoryear{{Renedo}, {Althaus}, {Miller Bertolami},
  {Romero}, {C{\'o}rsico}, {Rohrmann}  \& {Garc{\'{\i}}a-Berro}}{{Renedo}
  et~al.}{2010}]{2010ApJ...717..183R}
{Renedo} I.,  {Althaus} L.~G.,  {Miller Bertolami} M.~M.,  {Romero} A.~D.,
  {C{\'o}rsico} A.~H.,  {Rohrmann} R.~D.,   {Garc{\'{\i}}a-Berro} E.,  2010,
  \mn@doi [\apj] {10.1088/0004-637X/717/1/183}, \href
  {http://adsabs.harvard.edu/abs/2010ApJ...717..183R} {717, 183}

\bibitem[\protect\citeauthoryear{{Rolland}, {Bergeron}  \&
  {Fontaine}}{{Rolland} et~al.}{2018}]{2018ApJ...857...56R}
{Rolland} B.,  {Bergeron} P.,   {Fontaine} G.,  2018, \mn@doi [\apj]
  {10.3847/1538-4357/aab713}, \href
  {http://adsabs.harvard.edu/abs/2018ApJ...857...56R} {857, 56}

\bibitem[\protect\citeauthoryear{{Romero}, {C{\'o}rsico}, {Althaus}, {Kepler},
  {Castanheira}  \& {Miller Bertolami}}{{Romero}
  et~al.}{2012}]{2012MNRAS.420.1462R}
{Romero} A.~D.,  {C{\'o}rsico} A.~H.,  {Althaus} L.~G.,  {Kepler} S.~O.,
  {Castanheira} B.~G.,   {Miller Bertolami} M.~M.,  2012, \mn@doi [\mnras]
  {10.1111/j.1365-2966.2011.20134.x}, \href
  {http://adsabs.harvard.edu/abs/2012MNRAS.420.1462R} {420, 1462}

\bibitem[\protect\citeauthoryear{{Romero}, {Campos}  \& {Kepler}}{{Romero}
  et~al.}{2015}]{2015MNRAS.450.3708R}
{Romero} A.~D.,  {Campos} F.,   {Kepler} S.~O.,  2015, \mn@doi [\mnras]
  {10.1093/mnras/stv848}, \href
  {http://adsabs.harvard.edu/abs/2015MNRAS.450.3708R} {450, 3708}

\bibitem[\protect\citeauthoryear{{Schlegel}, {Finkbeiner}  \&
  {Davis}}{{Schlegel} et~al.}{1998}]{1998ApJ...500..525S}
{Schlegel} D.~J.,  {Finkbeiner} D.~P.,   {Davis} M.,  1998, \mn@doi [\apj]
  {10.1086/305772}, \href {http://adsabs.harvard.edu/abs/1998ApJ...500..525S}
  {500, 525}

\bibitem[\protect\citeauthoryear{{Schmidt}}{{Schmidt}}{1968}]{1968ApJ...151..393S}
{Schmidt} M.,  1968, \mn@doi [\apj] {10.1086/149446}, \href
  {http://adsabs.harvard.edu/abs/1968ApJ...151..393S} {151, 393}

\bibitem[\protect\citeauthoryear{{Schmidt}}{{Schmidt}}{1975}]{1975ApJ...202...22S}
{Schmidt} M.,  1975, \mn@doi [\apj] {10.1086/153949}, \href
  {http://adsabs.harvard.edu/abs/1975ApJ...202...22S} {202, 22}

\bibitem[\protect\citeauthoryear{{Sion}, {Greenstein}, {Landstreet}, {Liebert},
  {Shipman}  \& {Wegner}}{{Sion} et~al.}{1983}]{1983ApJ...269..253S}
{Sion} E.~M.,  {Greenstein} J.~L.,  {Landstreet} J.~D.,  {Liebert} J.,
  {Shipman} H.~L.,   {Wegner} G.~A.,  1983, \mn@doi [\apj] {10.1086/161036},
  \href {http://adsabs.harvard.edu/abs/1983ApJ...269..253S} {269, 253}

\bibitem[\protect\citeauthoryear{{Stobie}, {Ishida}  \& {Peacock}}{{Stobie}
  et~al.}{1989}]{1989MNRAS.238..709S}
{Stobie} R.~S.,  {Ishida} K.,   {Peacock} J.~A.,  1989, \mn@doi [\mnras]
  {10.1093/mnras/238.3.709}, \href
  {http://adsabs.harvard.edu/abs/1989MNRAS.238..709S} {238, 709}

\bibitem[\protect\citeauthoryear{{Torres}, {Garc{\'{\i}}a-Berro}, {Burkert}  \&
  {Isern}}{{Torres} et~al.}{2002}]{2002MNRAS.336..971T}
{Torres} S.,  {Garc{\'{\i}}a-Berro} E.,  {Burkert} A.,   {Isern} J.,  2002,
  \mn@doi [\mnras] {10.1046/j.1365-8711.2002.05830.x}, \href
  {http://adsabs.harvard.edu/abs/2002MNRAS.336..971T} {336, 971}

\bibitem[\protect\citeauthoryear{{Tremblay} \& {Bergeron}}{{Tremblay} \&
  {Bergeron}}{2008}]{2008ApJ...672.1144T}
{Tremblay} P.-E.,  {Bergeron} P.,  2008, \mn@doi [\apj] {10.1086/524134}, \href
  {http://adsabs.harvard.edu/abs/2008ApJ...672.1144T} {672, 1144}

\bibitem[\protect\citeauthoryear{{Tremblay} et~al.,}{{Tremblay}
  et~al.}{2017}]{2017MNRAS.465.2849T}
{Tremblay} P.-E.,  et~al., 2017, \mn@doi [\mnras] {10.1093/mnras/stw2854},
  \href {http://adsabs.harvard.edu/abs/2017MNRAS.465.2849T} {465, 2849}

\bibitem[\protect\citeauthoryear{{Winget} \& {Kepler}}{{Winget} \&
  {Kepler}}{2008}]{2008ARA&A..46..157W}
{Winget} D.~E.,  {Kepler} S.~O.,  2008, \mn@doi [\araa]
  {10.1146/annurev.astro.46.060407.145250}, \href
  {http://adsabs.harvard.edu/abs/2008ARA%26A..46..157W} {46, 157}

\bibitem[\protect\citeauthoryear{{Winget}, {Hansen}, {Liebert}, {van Horn},
  {Fontaine}, {Nather}, {Kepler}  \& {Lamb}}{{Winget}
  et~al.}{1987}]{1987ApJ...315L..77W}
{Winget} D.~E.,  {Hansen} C.~J.,  {Liebert} J.,  {van Horn} H.~M.,  {Fontaine}
  G.,  {Nather} R.~E.,  {Kepler} S.~O.,   {Lamb} D.~Q.,  1987, \mn@doi [\apjl]
  {10.1086/184864}, \href {http://adsabs.harvard.edu/abs/1987ApJ...315L..77W}
  {315, L77}

\bibitem[\protect\citeauthoryear{{Woosley} \& {Heger}}{{Woosley} \&
  {Heger}}{2015}]{2015ApJ...810...34W}
{Woosley} S.~E.,  {Heger} A.,  2015, \mn@doi [\apj]
  {10.1088/0004-637X/810/1/34}, \href
  {http://adsabs.harvard.edu/abs/2015ApJ...810...34W} {810, 34}

\bibitem[\protect\citeauthoryear{{Yuan}, {Liu}  \& {Xiang}}{{Yuan}
  et~al.}{2013}]{2013MNRAS.430.2188Y}
{Yuan} H.~B.,  {Liu} X.~W.,   {Xiang} M.~S.,  2013, \mn@doi [\mnras]
  {10.1093/mnras/stt039}, \href
  {http://adsabs.harvard.edu/abs/2013MNRAS.430.2188Y} {430, 2188}

\bibitem[\protect\citeauthoryear{{Yungelson}}{{Yungelson}}{2008}]{2008AstL...34..620Y}
{Yungelson} L.~R.,  2008, \mn@doi [Astronomy Letters]
  {10.1134/S1063773708090053}, \href
  {http://adsabs.harvard.edu/abs/2008AstL...34..620Y} {34, 620}

\makeatother
\end{thebibliography}







\bsp	
\label{lastpage}
\end{document}